\documentclass[usenatbib]{mnras}

\usepackage{newtxtext,newtxmath}
\usepackage{blindtext}
\usepackage[T1]{fontenc}
\usepackage{float}
\usepackage{cleveref}
\crefname{section}{§}{§§}
\Crefname{section}{§}{§§}

\DeclareRobustCommand{\VAN}[3]{#2}
\let\VANthebibliography\thebibliography
\def\thebibliography{\DeclareRobustCommand{\VAN}[3]{##3}\VANthebibliography}

\usepackage{graphicx}
\usepackage{amsmath}

\newcommand{\pcmsq}{cm$^{-2}$}

\newcommand{\civ}{\ion{C}{iv}}
\newcommand{\hi}{\ion{H}{i}}
\newcommand{\Ociv}{$\Omega_{\rm CIV}$ }

\defcitealias{Puchwein2015}{P15}
\defcitealias{Kartick2021}{S21}
\defcitealias{HaardtMadau2012}{HM12}

\usepackage{soul}

\title[Tracking ionization in IGM]{Fast modelling of ionization balance in the intergalactic medium: I- implications for the IGM metallicity}

\author[Arya et al.]{
Bhaskar Arya,$^{1,2}$\thanks{bhaskararya4@gmail.com},
Kartick C. Sarkar$^{1,2}$\thanks{kcsarkar@rrimail.rri.res.in}, Shiv K. Sethi$^2$
\\
$^{1}$ Dept of Space, Planetary and Astronomical Sciences and Engineering (SPASE), Indian Institute of Technology, Kanpur, India\\
$^{2}$Raman Research Institute, Bengaluru, India\\
}

\date{Accepted XXX. Received YYY; in original form ZZZ}
 
\pubyear{\the\year{}}

\begin{document}
\label{firstpage}
\pagerange{\pageref{firstpage}--\pageref{lastpage}}
\maketitle

\begin{abstract}
Ionization balance in the intergalactic medium (IGM) underpins the interpretation of quasar absorption spectra by linking observed ionic columns to the underlying gas density, temperature, metallicity, and ionizing radiation field. Yet the relevant microphysical timescales (ionization, recombination, and cooling) can be comparable to the timescales over which the ultraviolet background (UVB) and thermodynamic state evolve. As a result, ion populations may exhibit strong history dependence. We present a fast, metals-inclusive, zero-dimensional framework to evolve the IGM across redshifts. The framework follows a coupled thermal and ionization evolution of a Lagrangian gas parcel in a redshift-dependent UVB. The model integrates stiff, time-dependent rate equations for H, He, and 107 metal ions (C, N, O, Ne, Mg, Si, S, Fe) while self-consistently evolving the thermodynamics via photoheating and standard cooling channels. We validate the approach by comparing its predictions to full three-dimensional hydrodynamical non-equilibrium calculations, finding that it recovers the IGM thermal and ionization histories to good accuracy over a wide redshift range, including the characteristic heating associated with He\,\textsc{ii} reionization. 
The framework efficiently maps ionization histories into cosmic metal-ion abundances and ionization corrections relevant for metallicity inference. As an application, we predict the cosmic C\,\textsc{iv} density parameter, $\Omega_{\rm CIV}$, and use it to infer the origin of metal ions in the IGM and IGM metallicities from observational measurements, obtaining values in reasonable agreement with literature constraints. We also discuss the possible implications for the non-equilibrium ionization (NEI) in the IGM. Our framework is designed for rapid parameter studies, enabling controlled tests of how reionization timing, UVB spectral hardness, self-shielding, and spatial UVB inhomogeneity (e.g., patchy reionization and local sources) propagate into the IGM thermal/ionization history and metal-line observables, providing a practical bridge between micro-physical modeling and absorption-line analyses.
\end{abstract}

\begin{keywords}
intergalactic medium -- metal absorption -- simulations
\end{keywords}

\section{Introduction}

The thermal evolution of the intergalactic medium (IGM) is governed by a complex interplay of heating and cooling processes. The dominant mechanisms shaping the temperature-density relation of the IGM include photoheating from the metagalactic ultraviolet background (UVB), balanced against adiabatic cooling due to cosmic expansion. Additional processes, such as shock heating, compressional heating, and radiative cooling, also contribute, particularly at higher densities. A key characteristic of the IGM's thermal state is its mean temperature and the power law temperature-density relation, which encodes the cumulative impact of these physical processes at moderate overdensities ($\delta \lesssim 10$) \citep{HuiGnedin1997, Theuns1998, Schaye_2000,Furlanetto_2009, Puchwein2015, Sanderbeck_2016, McQuinn_2016}.

Understanding this thermal history is essential not only for theoretical models of cosmic reionization and structure formation but also for interpreting observational tracers such as the Ly$\alpha$ forest. The Ly$\alpha$ forest, a series of \ion{H}{i} absorption features in the spectra of distant quasars, serves as a sensitive probe of the IGM at low to moderate over-densities \citep[$\delta \sim 10;$][]{Croft_1998, Rauch_1998, Hui_1999, Weinberg_2003, Meiksin_2009, McQuinn_2016}. However, accurate interpretation of such data relies heavily on precise modeling of the underlying thermal state of the IGM.

A major source of uncertainty in current models arises from the treatment of radiative transfer during the epoch of reionization (EoR) \citep{Iliev_2006, Tittley_2007, McQuinn_2009, Faucher_2009, GnedinMadau2022}. Most cosmological hydrodynamical simulations assume photoionization equilibrium, thereby neglecting transient non-equilibrium effects that can significantly alter the thermal history. These effects are particularly important during reionization, when rapid changes in ionization state can lead to excess heating that equilibrium models fail to capture. To compensate, simulations often artificially boost photo-heating rates to match observational constraints, but such approximations may not capture the full physical picture \citep{Schaye_2000, Puchwein2015, Bolton_2016, Sanderbeck_2016, Gaikwad_2017}.

From the observational side, metallicities and densities in IGM absorbers are usually inferred with single-zone \texttt{CLOUDY} models under photo-ionization equilibrium (PIE) and a spatially uniform UV background (typically \citet{HaardtMadau2012}; hereafter \citetalias{HaardtMadau2012}), which provide the ionization corrections applied to measured column densities \citep{Aguirre_2002, Schaye_2003, D_Odorico_2010, Simcoe_2011, Werk_2014, Fumagalli_2016, Lehner_2016}. This approach is standard for Lyman-limit and partial Lyman-limit systems and for low-$z$ CGM analyses (e.g., COS-Halos), where suites of low/high ions (C\,\textsc{ii/iii/iv}, Si\,\textsc{ii/iii/iv}, O\,\textsc{vi}) are fit in PIE to recover metallicity ($Z$), ionization parameter ($U$), and hydrogen number density ($n_{\mathrm H}$) \citep{Lehner_2013, Werk_2014, Fumagalli_2016, Wotta_2019, Ferland_2017}. However, PIE can misestimate ion fractions and thus bias $Z$, $n_{\mathrm H}$, and $T$ when gas undergoes rapid heating/cooling, shocks, or time-variable illumination; non-equilibrium calculations for cooling gas and fluctuating radiation fields demonstrate history dependent departures from equilibrium \citep{Gnat_2007, Vasiliev_2011, Oppenheimer_2013, Richlings_2014a, Richlings_2014b}. E.g., flickering AGN produces long-lived “proximity-zone fossils” in which high ions (notably \ion{O}{vi}) remain over-ionized for $\sim 10^{7}$ yr, boosting column densities by factors of a few and leading to order-unity discrepancies if analyzed with PIE \citep{Segers_2017, Oppenheimer_2017}. Given today’s high-S/N, multi-ion spectra and precise $N_{\mathrm{H\,I}}$ measurements from Lyman-limit coverage, ionization model systematics now dominate the error budget. Therefore, testing non-equilibrium ionization (NEI) (or isolating phases where PIE is valid) is required for robust metallicity inferences \citep{McQuinn_2016}.

On the simulation side, these concerns have motivated fully time-dependent, self-consistent treatments: cosmological SPH runs that evolve the H/He ionization state via rate equations under an evolving, homogeneous UVB capture the excess heating and thermal "memory" that PIE-based heating boosts cannot reproduce. Notably, \citet{Puchwein2015} (hereafter \citetalias{Puchwein2015}) implemented an NEI ionization module on \texttt{P-GADGET3} simulation and tracked H \& He (with metals excluded) all the way from $z \sim 16$ to the present day. This more physically motivated treatment allows for a more accurate representation of the IGM’s thermal history as well as ionization structure, enabling improved interpretation of Ly$\alpha$ forest statistics and other IGM observables. Relative to equilibrium treatments, \citetalias{Puchwein2015} predicts a systematically higher temperature at mean density, $T_0$, during reionization and a delayed response of the gas to ionizing radiation, allowing for realistic heat injection due to reionization fronts. This can amount to differences of order a few $\times 10^3$K ($\sim 40\%$) in mean-density temperature at $z \sim 3-5$, shifting the inferred $T-\Delta$ parameters ($T_0$, $\gamma_{\textrm{fit}}$)\footnote{We distinguish the fit parameter $\gamma_{\textrm{fit}}$ from the conventional adiabatic index $\gamma$ (= 5/3 for monoatomic gas).} in ways that impact the Ly$\alpha$ line widths and flux power spectrum inferences. Additionally, it also leads to more reliable estimates of heating and cooling rates, since they are computed from the correct ionization state of the gas.

\citealt{Oppenheimer_2016}, building on \citealt{Oppenheimer_2013}, evolved a metals-inclusive, time-dependent ionization/cooling network ($\sim$136 ions of 11 elements) coupled to the hydrodynamics from $z < 0.5$ under an evolving metagalactic UVB, thereby avoiding PIE post-processing. They demonstrate that non-equilibrium ionization lags in cooling or shock-heated CGM gas, substantially shifting the predicted abundances of high ions (e.g., \ion{O}{vi}, \ion{N}{v}, and \ion{Ne}{viii}) relative to PIE, yielding more self-consistent low-$z$ column density distributions, covering fractions, and line-width statistics, and linking high-ion phases to recent feedback, mixing, and cooling flows (including the persistence of over-ionization after transient heating). The work quantifies systematic biases that arise when inferring $Z$, $U$, $n_{\rm H}$, and $T$ with PIE where NEI effects are non-negligible.

Notwithstanding the benefits, 3D cosmological simulations incorporating metals-inclusive, non-equilibrium ionization and thermal evolution entail significant computational expense. This high cost arises from the need to resolve both spatial structure and time-dependent microphysical processes across large dynamic ranges. Consequently, \citetalias{Puchwein2015} ignored the self-consistent evolution of metals, which must be treated via equilibrium post-processing, limiting predictive power for CGM/LLS metal absorbers. On the other hand, in \citet{Oppenheimer_2016} the NEI metals module is typically activated only at late times (\(z\lesssim0.5\)), so time-dependent metal ionization is not followed through earlier epochs. As a result, comprehensive scans over reionization histories, UVB models, and auxiliary heating mechanisms become intractable at scale. The need for multiple realizations and high-resolution runs further amplifies the cost, impeding rigorous tests of new physics.

To mitigate these challenges, we develop a zero-dimensional (0D) simulation framework that evolves a single Lagrangian gas parcel over time \citep[based on the NEI package of][]{Kartick2021}; hereafter \citetalias{Kartick2021}, as opposed to a full 3D hydrodynamic simulation. The approach uses redshift-dependent UVB and captures the full range of relevant heating and cooling mechanisms, such as photoheating, adiabatic expansion, gravitational compression/decompression, inverse Compton cooling, and radiative cooling, while significantly reducing the computational cost. We track the thermal history of the gas densities as well as the abundance and ionization states of hydrogen and helium, and ions of carbon in the IGM by solving the time-dependent rate equations. \footnote{Although, in principle we can also track other relevant metals (N, O, Ne, Mg, Si, S, Fe).} By focusing on the thermal and ionization evolution of a single gas parcel without spatial complexity, this controlled setup allows for efficient exploration of a wide parameter space (e.g., reionization epoch, UV background, etc.) and facilitates the inclusion of additional physical processes (e.g., self-shielding, dust heating, local radiation field etc.) that would be computationally infeasible to test in full 3D simulations.

It also enables controlled, repeatable experiments that we can validate against NEI hydrodynamic runs and analytic expectations, providing a baseline for interpreting full 3D results. E.g., in most cosmological hydrodynamics codes, metallicity is carried as an advected scalar (often element-by-element) set by subgrid enrichment: newly formed stars return metals according to stellar yield tables and feedback prescriptions; the metals are then transported by galactic winds and mixed numerically or via explicit diffusion models \citep{Wiersma_2009a, Wiersma_2009b, Shen_2010, Vogelsberger_2013, Sarkar2015, Schaye_2015, Pillepich_2018, Hopkins_2018}. Because simulations track total metal mass rather than ion states, comparisons to data typically convert $(Z, n_{\rm H}, T)$ into observables through equilibrium ionization post-processing with a chosen UV background (e.g. \texttt{CLOUDY} + \citetalias{HaardtMadau2012}), sometimes adding metallicity floors or tuning feedback to match global relations such as the galaxy mass-metallicity relation \citep{HaardtMadau2012, Torrey_2012, Ferland_2017, Hummels_2017}. Our metals-inclusive, one-zone NEI framework helps refine this pipeline by mapping a parcel’s density/temperature/UVB history to time-dependent ion fractions and emissivities, thereby quantifying biases that arise when PIE is assumed in non-equilibrium regimes \citep{Oppenheimer_2013, Richlings_2014a, Richlings_2014b, Oppenheimer_2016}. In practice, it can allow (i) NEI corrections to infer $Z$ from multi-ion columns, (ii) benchmarking metallicity in hydrodynamic simulations, and (iii) generation of NEI lookup tables/emulators for snapshot post-processing across plausible thermodynamic histories.

In this work, we demonstrate the robustness of our approach by (i) recovering the IGM thermal and ionization history in close agreement with state-of-the-art 3D NEI hydrodynamical simulations, and (ii) deriving metallicities from C\,\textsc{iv} absorbers. \cref{sec:0D} introduces the zero-dimensional NEI model and details all the physical processes included in our framework. \cref{sec:res} presents results: the temperature-density parameters and their sensitivity to density, UVB, reionization redshift, and self-shielding, along with C\,\textsc{iv} predictions and the resulting metallicity estimates. Further, we discuss the implications of these results in \cref{sec:discussions}. Finally, in \cref{sec:conclusion}, we conclude with a summary and implications for applying NEI ionization modeling to IGM inference.

%%%%%%%%%%%%%%%%%%%%%%%%%%%%%%%%%%%%%%%%%%%%%%
\section{The 0-D framework}
\label{sec:0D}
%%%% 
In this section, we model the ionization and heating exclusively from the metagalactic ultraviolet background (UVB) within a spatially uniform approximation. The radiation field is represented by the angle-averaged specific intensity $J_\nu (z, t)$ from the cosmological UVB model \citetalias{HaardtMadau2012} throughout this work, without the contributions from local sources or geometric radiation transport. The C\,\textsc{iv} absorption in our modeling arises predominantly from moderately over-dense environments, with characteristic over-densities \(\Delta \sim 10\) and $T \sim \textrm{few} \times 10^4$ K. In this regime, neutral hydrogen columns can be sufficiently large that the gas is no longer guaranteed to be optically thin at the Lyman edge, and H\,\textsc{i} self-shielding can significantly attenuate the incident metagalactic UVB. Because the C\,\textsc{iv}-bearing gas may therefore experience a reduced ionizing flux relative to the uniform-background limit, we explicitly bracket the UVB treatment with two cases.
(i) \emph{Optically thin UVB:} as a fiducial case (and unless stated otherwise), we adopt the spatially uniform, optically thin UVB (\citetalias{HaardtMadau2012}) without attenuation. (ii) \emph{Self-shielded UVB:} a case variation where we model local attenuation by neutral hydrogen via an effective shielding factor. We discuss this in more detail in \cref{subsec:var}. These two limiting cases provide a controlled estimate of the sensitivity of the inferred C\,\textsc{iv} ionization state (and derived \(\Omega_{\mathrm{C\,IV}}\)) to self-shielding in the over-dense gas that dominates the signal.

Photoionization by the metagalactic UVB sets the ion fractions and provides a source term for thermal energy via photo-heating. The gas temperature evolution is governed by adiabatic processes (Hubble cooling and localized expansion/compression in over-densities) plus non-adiabatic heating and cooling terms, including UVB photo-heating and radiative losses (and inverse Compton cooling where relevant). We evolve a full H/He+metals ion network; however, in the thermal energy budget, we include photo-heating only from H\,\textsc{i}, He\,\textsc{i}, and He\,\textsc{ii}, and neglect metal photo-heating in this work. We have tested the photo-heating from metals in some cases and found it to be negligible. Shock heating is noticeable around massive halos and is dependent on the mass of the halo and distance from it, a feat not achievable by 0-D evolution. Therefore, we exclude modeling shock heating at the expense of limiting the applicability of our results to only the IGM regions ($\delta \lesssim 50$). Additionally, this omission is appropriate for our application, as the observed Doppler parameters of C\,\textsc{iv} absorbers ($b_{\mathrm{CIV}}\!\sim\!10~\mathrm{km\,s^{-1}}$; \citealt{Lehner_2016}) indicate that the bulk of the C\,\textsc{iv} population resides in low temperature photoionized gas,  \citep{Kim_2013}.

\subsection{Photoionization and photo-heating from the UVB}
\label{subsec:photo-heating}
%%%%%
The hydrogen and helium photoionization rates (per ion number) are given by,
\begin{equation}
    \Gamma_i = \int_{\nu_i}^{\infty}\textrm{d}\nu \frac{4\pi J_{\nu}}{h\nu}\sigma_i(\nu)
\end{equation}
where $h$ is the Planck constant, $\nu$ is the frequency, $J_{\nu}$ is the space- and angle-averaged monochromatic intensity, the subscript $i$ denotes the relevant ion species, $h\nu_i$ is the ionization energy, and $\sigma_i(\nu)$ is the photoionization cross-section. Therefore, the spatially uniform photoheating rate (per ion number) is given by,
\begin{equation}
    \mathcal{H}_i = \int_{\nu_i}^{\infty}\textrm{d}\nu \frac{4\pi J_{\nu}}{h\nu}h(\nu - \nu_i)\sigma_i(\nu)
\end{equation}
The total photoheating rate per unit volume is then given by,
\begin{equation}
    \mathcal{H} = \sum_i n_i\mathcal{H}_i
    \label{eq:heating}
\end{equation}
where $n_i$ is the number density of each ion species. Since we do not consider heating from metals in this work,  $i \equiv (\textrm{H\,\textsc{i}, He\,\textsc{i}, He\,\textsc{ii}})$.

\subsection{Evolution of pressure}
\label{subsec:evo-pressure}
%%%%%%
In contrast to the standard approach in literature \citep{HuiGnedin1997, Sanderbeck_2016, McQuinn_2016}, which evolves the temperature of a Lagrangian gas element, we instead use the pressure, $P$, as the primary evolution variable.
\begin{equation}
    \frac{dP}{dt} = -3\gamma HP + \frac{\gamma P}{1+\delta}\frac{d\delta}{dt} + (\gamma-1)\frac{dQ}{dt}
    \label{eq:prs_K24}
\end{equation}
where $H$ is the Hubble parameter, $\delta$ is the density contrast, $k_{\rm B}$ is the Boltzmann constant, $\gamma$ is the adiabatic index ($=5/3$ for a monatomic ideal gas), and $Q$ denotes the net energy per unit volume where,
\begin{equation}
    \frac{dQ}{dt} = \left.\frac{dQ}{dt}\right|_{\rm ph} + \left.\frac{dQ}{dt}\right|_{\rm ct} + \left.\frac{dQ}{dt}\right|_{\rm rc} + \left.\frac{dQ}{dt}\right|_{\rm ic}
    \label{eq:dQdt}
\end{equation}
The first and second terms on the right-hand side of Eq.~\ref{eq:prs_K24} describe adiabatic effects: cooling from the Hubble expansion and heating (or cooling) from local gravitational compression (or decompression), respectively. The third and final term aggregates all non-adiabatic sources and sinks, as detailed in Eq.~\ref{eq:dQdt}. The first two terms herein provide heating from photoheating and charge transfer (with photoheating dominating at all redshifts), the third term corresponds to loss via radiative cooling calculated based on the local non-equilibrium ion density following \citealt{Gnat_2012}. The final term represents cooling off inverse Compton scattering as given in \citet{Seager_2000}
\begin{equation}
    \left.\frac{dT}{dt}\right|_{\rm ic} = \frac{(\gamma - 1)}{n_{\rm b}k_{\rm B}}\left.\frac{dQ}{dt}\right|_{\rm ic} = -\frac{8\sigma_{\rm T}Un_{\rm e}}{3m_{\rm e}cn_{\rm b}}(T-T_{\rm CMB})
\end{equation}
where $T$ is temperature of the gas, $T_{\rm CMB}$ is temperature of the cosmic microwave background (CMB), $\sigma_{\rm T}$ is the Thomson scattering cross-section, $U=a_{\rm R}T_{\rm CMB}^4$ is the radiation energy density with $a_{\rm R} = \frac{8\pi^5k_{\rm B}^4}{15h^3c^3}$ being the radiation constant, $h$ is Planck's constant, $m_{\rm e}$ is electron mass, and $c$ is speed of light. The inverse Compton is relevant at $z \gtrsim 10$ during reionization of H+He therefore, one can safely assume hydrogen is completely ionized and helium is singly ionized, $n_{\rm b} \approx 2n_{\rm e}$, and $T \gg T_{\rm CMB}$.The temperature is then calculated under the assumption of ideal gas, $T = \frac{P}{n_{\textrm{b}}k_{\textrm{B}}} = \frac{P}{\rho} \frac{\mu m_{\rm p}}{k_{\rm B}}$, where $\mu$ is the mean molecular weight and is calculated based on the local ionization fractions of H, He, and other metals.

\subsection{Evolution of density}
Evolution of the density is represented by the $d\delta/dt$ term in Eq.~\ref{eq:prs_K24}. Since there are no particles in our framework, we evolve the Lagrangian baryon overdensity directly as a continuous variable rather than following mass elements in an $N$-body sense. We implement two complementary prescriptions for the density trajectory: (i) first-order Lagrangian perturbation theory (1LPT), where the overdensity follows the linear growth factor, (ii) spherical collapse model (SCM), using the standard top-hat evolution to capture nonlinear compression/expansion. While these formalisms are classically applied to dark matter, we use them for baryons as a controlled approximation at the parcel level. On sufficiently large scales, the baryon field tracks the dark matter, and deviations at small scales (e.g., pressure/Jeans smoothing and delayed collapse) are ignored in our thermal/pressure terms. This setup lets us prescribe plausible density histories for the gas parcel and cleanly isolate the impact of heating/cooling physics.

We begin with the linear-theory description of density evolution, in which the density contrast grows as
\begin{equation}
    \delta(a) = \frac{D(a)}{D(a_i)}\delta(a_i)
\end{equation}
where $a$ is the scale factor, $a_i$ is the initial scale factor and $D(a)$ is the linear growth factor given by
\begin{equation}
    D(a) = \frac{5}{2}\Omega_{\rm m, 0}E(a)\int_0^a\frac{da^{\prime}}{(a^{\prime}E(a^{\prime}))^3}
\end{equation}
with
\begin{equation}
    E(a) = \sqrt{\Omega_{\rm m, 0}/a^3 + \Omega_{\Lambda,0}}.
\end{equation}
Under these assumptions, the rate of growth of the density contrast is
\begin{equation}
    \dot{\delta} = \frac{\Omega_{\rm m, 0}H_0}{2a^2E(a)}\left(5 - 3\frac{D(a)}{a}\right)\frac{\delta(a_i)}{D(a_i)}.
\end{equation}

Henceforth, we refer to this linear density evolution as the \textit{fiducial} case because of its simplicity, its ability to reproduce the large-scale density field, and its avoidance of shell crossing. We discuss these advantages further in \cref{sec:res}.

For comparison, we also consider the non-linear spherical collapse model for the evolution of the density contrast, following \citet{Fosalba_1998, Mandal_2020}:
\begin{equation}
    \ddot{\delta} + 2H\dot{\delta} - \frac{4\dot{\delta}^2}{3(1+\delta)} - \frac{3}{2}H^2\Omega_{\textrm{m}}(a)\delta(1+\delta) = 0.
\end{equation}

\subsection{Initial conditions}
We initialize the calculation at \(z_i=19\) (\(a_i=0.05\)), and assume the gas is fully neutral. The initial temperature is set to the CMB temperature at that epoch,
\(T_i = T_{\rm CMB,0}/a_i = 2.725~\mathrm{K}/0.05 = 54.5~\mathrm{K}\), which fixes the initial pressure. The metagalactic UV background is switched on at \(z\simeq 15.1\) following \citetalias{HaardtMadau2012}, and metals are introduced at \(z=7\). We adopt a constant metallicity ($\log Z/Z_{\odot}=-3$) at all redshifts, although a redshift-dependent prescription can be incorporated straightforwardly. We have tested that increasing the metallicity of the IGM to even $\log Z/Z_{\odot}= -1$ does not affect radiative cooling or photo-heating. For consistency with \citetalias{Puchwein2015}, we adopt a flat \(\Lambda\)CDM cosmology with \(\Omega_{\rm m}=0.305\), \(\Omega_\Lambda=0.695\), \(\Omega_{\rm b}=0.0481\), and \(H_0=67.9\). We solve the evolution of 45 overdensity points uniformly distributed over a range of $\delta \sim -0.05$ to $60$ at $z=19$.

\subsection{Ionization network}
\label{sec:ion_network}
We evolve the ionization state of the gas by following the ion fractions $x_{k,i}$ of ionization stage $i$ of element $k$ in a Lagrangian gas parcel, as described in \citet{Gnat_2007, Kartick2021}. Since our framework is zero-dimensional, it does not include any spatial advection term. Therefore, the ion fractions satisfy
\begin{equation}
\frac{dx_{k,i}}{dt} = S_{k,i},
\end{equation}
where $S_{k,i}$ is the net source term for ion $i$. For a given element, the source term $S_{k,i}$ may be written schematically as
\begin{equation}
\begin{split}
S_{k,i}
={}&
n_e \left[
x_{k,i+1}\,\alpha_{k,i+1}
- x_{k,i}\left(\xi_{k,i}+\alpha_{k,i}\right)
+ x_{k,i-1}\,\xi_{k,i-1}
\right] \\
&- x_{k,i}\,\Gamma_{k,i} + A_{k,i},
\end{split}
\end{equation}
Here, $\alpha_{k,i}$ denotes the total recombination coefficient from ion $(k,i)$ to $(k,i-1)$, $\xi_{k,i}$ is the collisional ionization rate coefficient from ion $(k,i)$ to $(k,i+1)$, $\Gamma_{k,i}$ is the photoionization rate, $A_{k,i}$ is the Auger ionization rate contribution from lower ionization states into the current ion, and $n_e$ is the electron number density. In our calculation, the total recombination coefficient includes both radiative and dielectronic recombination, while the ionization and recombination rates also include charge-transfer contributions. We refer the reader to \citetalias{Kartick2021} for further details.

The evolution of each ion is therefore coupled to neighbouring ionization states of the same element, while the full network is further coupled through the electron density and temperature.
We integrate the coupled ionization-thermal equations using an adaptive Cash-Karp method of order 4(5) \citep{CashKarp1990}, which adjusts the timestep to resolve rapid changes in the ion fractions and temperature. At each step, the electron density is updated self-consistently from the instantaneous ion fractions, and we monitor the normalization condition
\begin{equation}
\sum_i X_{k,i} = 1
\end{equation}
for every element to ensure abundance conservation to numerical accuracy. Although the framework is designed to evolve a broad metal-ion network, in the present work we restrict the metal evolution to carbon only, and defer the full metal-ion network to future work.

%%%%%%%%%%%%%%%%%%%%%%%%%%%%%%%%%%%%%%%%%%%%%
\section{Results}
\label{sec:res}
\subsection{Recovery of thermal and ionization history}

\begin{figure}
\centering
\includegraphics[width=\columnwidth]{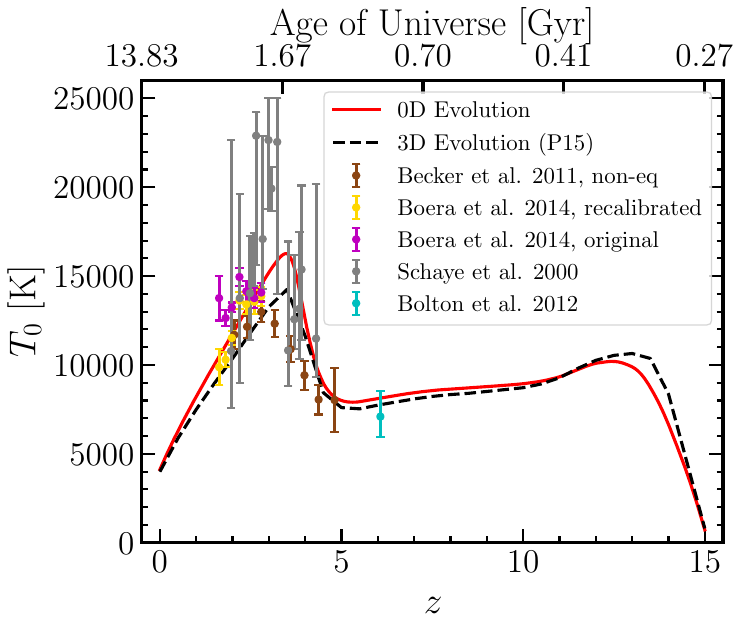}
\caption{Evolution of $T_0$ \textit{w.r.t.} redshift for our zero dimension NEI simulation (shown in red curve) compared to full 3D NEI hydro simulation (as taken from \citetalias{Puchwein2015}; shown in black dashed line). Overlaid are observational constraints on $T_0$ as shown in colored circles \citep{Schaye_2000, Becker_2011, Bolton_2012, Boera_2014}. We find that the zero-dimensional model reproduces the 3D NEI thermal history to within a few $\times 10^3$ K ($\sim 15\%$) over the entire cosmic time, including the timing of the He\,\textsc{ii} reionization bump.}
\label{fig:T0}
\end{figure}

In Fig.~\ref {fig:T0}, we present our inferred $T_0$, i.e., temperature at mean IGM density, as a function of redshift from the zero-dimensional (0D) model (red solid) and compare it with full 3D NEI simulations (black dashed). We additionally overlay observational constraints on $T_0$ (colored circles). 
We find close agreement in \(T_0(z)\) across redshift, with both simulations tracing the same thermal history. Temperature rises sharply when the UVB is switched on, driven by H\,\textsc{i} and He\,\textsc{i} reionization, and then enters a broad plateau as hydrogen becomes ionized and helium singly ionized. During this phase, the diffuse IGM is well described by photoionization equilibrium (expansion cooling vs photo-heating), until the UVB hardens sufficiently to ionize He\,\textsc{ii}, producing a second, more prominent heating peak at \(z\sim3.5\). Once the gas is fully ionized, the temperature declines gradually toward lower \(z\) owing to the dominance of the Hubble expansion. Our recovered temperatures are fully consistent with the full 3D NEI simulations within the observational $T-\Delta$ scatter, indicating that the simplified 0D model captures the dominant heating/cooling balance seen in full 3D NEI simulations. Modest deviations appear near the two reionization peaks, but the overall normalization and timing of features are well matched.

\begin{figure}
\centering
\includegraphics[width=\columnwidth]{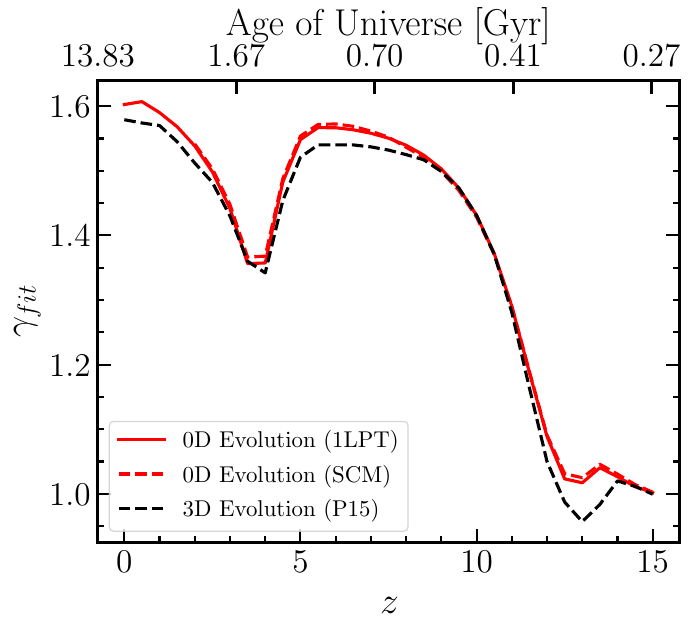}
\caption{Redshift evolution of the fitted IGM equation of state slope, \(\gamma_{\textrm{fit}}\), inferred from the \(T\)–\(\Delta\) relation in our zero-dimensional (0D) framework for the two density-growth prescriptions: linear theory (red solid) and, spherical collapse (red dashed). For both cases, we fit \(T(\Delta)=T_0\,\Delta^{\,\gamma_{\textrm{fit}}-1}\) over \(10^{-0.5} \leq \Delta \leq 2\). We compare our results to full 3D hydro simulations as reported in \citetalias{Puchwein2015} (black dashed). We see that the spherical collapse prescription typically yields a closer match in both normalization and timing at most redshifts.}
\label{fig:gamma}
\end{figure}

Our framework also recovers the temperature-density relation of the diffuse IGM, which is generally modeled as
\begin{equation}
T(\Delta,z) = T_0(z)\,\Delta^{\,\gamma_{\textrm{fit}}(z)-1},
\label{eq:TDelta}
\end{equation}
where \(\Delta \equiv (1 + \delta) = \rho_b/\bar{\rho}_b\) is the baryon overdensity, and \(\gamma_{\textrm{fit}}\) encodes the slope set by the balance of photoheating and cooling. The pair \((T_0,\gamma_{\textrm{fit}})\) captures the IGM’s recent heating history and pressure smoothing, and directly informs Ly\(\alpha\) observables such as line widths and the small-scale flux power. 
Figure \ref{fig:gamma} shows the recovered equation of state index, $\gamma_{\textrm{fit}}$, for which we evolve an ensemble of density trajectories using 1LPT and SCM, assuming only dark matter.  At each timestep we record the corresponding \((\Delta, T)\) pairs in the diffuse IGM limit (\(10^{-0.5} \leq \Delta \leq 2\)), and fit the power-law relation $T(\Delta) = T_0\,(\Delta)^{\,\gamma_{\textrm{fit}}-1}$ to infer \(\gamma_{\textrm{fit}}\). We compare these results with those obtained using full 3D hydrodynamical simulations in \citetalias{Puchwein2015}.

Both 0D curves track near identical thermal history: \(\gamma_{\textrm{fit}}\) rises from near-isothermal values at high \(z\) (indicating heating), flattens during the H\,\textsc{i}/He\,\textsc{i} reionization plateau, and increases again toward \(\gamma_{\textrm{fit}}\!\sim\!1.4\text{-}1.6\) as the UVB hardens and pressure smoothing strengthens around the He\,\textsc{ii} epoch. The large dip at $z\sim 4$ is due to the \ion{He}{ii} reionization when a part of the photo- and compressional-heating is going toward ionizing \ion{He}{ii} and as a result rise in temperature slows down. Our results agree remarkably well with the full 3D hydrodynamical simulations, with deviations of $\lesssim 2\%$ over most of cosmic time. This indicates that the choice of density trajectory primarily adjusts small residual differences rather than altering the overall evolutionary trend \citep{McQuinn_2015}. It is worth noting that the inferred $\gamma_{\mathrm{fit}}$ can depend on the density interval over which the fit is performed, with this sensitivity becoming more pronounced at higher redshifts ($z \gtrsim 5$). We also caution that a direct comparison with \citetalias{Puchwein2015} is not entirely straightforward. In the paper, the authors determine the local slope, $\gamma - 1$, at overdensity $\Delta$ using an iterative procedure (as described in Appendix E of \citetalias{Puchwein2015}), whereas in our analysis, we instead infer a density-independent slope by fitting a single power-law relation over a finite density range.

\begin{figure}
\centering
\includegraphics[width=\columnwidth]{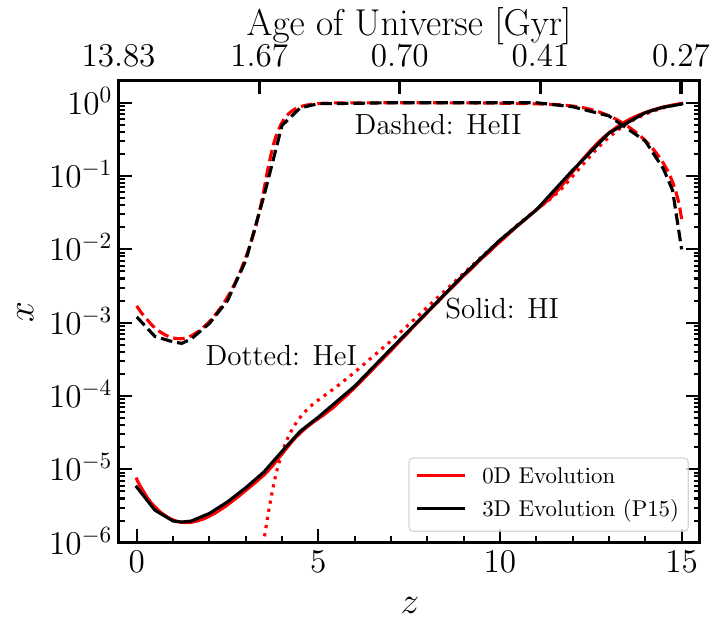}
\caption{Evolution of ion fractions of H\,\textsc{i} and He\,\textsc{ii} \textit{w.r.t.} redshift at mean density, for our zero dimension NEI simulation (shown in red curves) compared to full 3D NEI hydro simulation (as taken from \citetalias{Puchwein2015}; shown in black curves). We also show He\,\textsc{i} fraction for our model (dotted red). We again see that the zero-dimensional model recovers the ion fractions in very good agreement with 3D NEI simulations.}
\label{fig:ion_frac}
\end{figure}

In fig.~\ref{fig:ion_frac}, we compare the redshift evolution of neutral hydrogen, H\,\textsc{i}, and singly ionized helium, He\,\textsc{ii}, ion fractions. The solid curves show the H\,\textsc{i} fraction, \(x_{\mathrm{H\,I}}\), and the dashed curves show He\,\textsc{ii}, \(x_{\mathrm{He\,II}}\), with results from the 0D model and the 3D NEI simulation overlaid. For completeness, we also show the neutral helium fraction, \(x_{\mathrm{He\,I}}\), for our 0D model; this quantity is not available in the 3D NEI simulation and is therefore only plotted for the 0D case. At high redshift, \(x_{\mathrm{H\,I}}\!\approx\!1\) until the UVB turns on, after which it drops sharply and settles at \(x_{\mathrm{H\,I}}\sim10^{-5}\text{--}10^{-6}\). In contrast, \(x_{\mathrm{He\,II}}\) climbs to \(\sim1\) following the first reionization phase (H\,\textsc{i}/He\,\textsc{i}), remains elevated through the plateau, and then declines rapidly at \(z\simeq3\text{--}4\) during He\,\textsc{ii} reionization. The 0D (red) and 3D (black) curves are in close agreement for both species: the onset and timing of the H\,\textsc{i} drop and the He\,\textsc{ii} decline align well, with only minor, localized offsets near the transition edges, indicating that the simplified one-zone treatment reproduces the key ionization history captured by the full simulations.

\subsection{Prediction of \ion{C}{iv} abundance and metallicity}
As an example use of our framework, in this section, we estimate the \ion{C}{iv} abundance in the IGM and provide implications for the IGM metallicity. Since our 0D runs evolve individual Lagrangian gas parcels, one trajectory cannot reproduce the observed cosmic C\,\textsc{iv} abundance, which is an integral over absorbers sampling a wide range of column densities \citep{Songaila_2001, Pettini_2003, Simcoe_2004, Cooksey_2010, Simcoe_2011, D_Odorico_2010, Davies_2023, Yu_2025}
\begin{equation}
    \Omega_{\textrm{CIV}}(z) \equiv \frac{\rho_{\textrm{CIV}}(z)}{\rho_{\textrm{crit, 0}}} = \frac{H_0 m_{\textrm{C}}}{c\rho_{\textrm{crit, 0}}}\int N(z)\:f(N,z)\: dN
\end{equation}
where $f(N,z)$ is the column density distribution function of C\,\textsc{iv}, $\rho_{\textrm{CIV}}$ is the mass density of C\,\textsc{iv} in comoving units, and $\rho_{\textrm{crit, 0}}$ is the critical density of the Universe at $z = 0$.
To emulate a realistic IGM density field while retaining computational efficiency, we therefore run an ensemble of 0D simulations for \(N_\Delta=45\) representative overdensities \(\{\Delta_j\}\) spanning the diffuse and non-linear regime. For every \(\Delta_j\), we evolve the full NEI thermochemistry and record \(x_{\mathrm{C\,IV}}(\Delta_j,z)\). 
We then assign each density a redshift-dependent weight \(w_j(z)\) proportional to the baryon density probability distribution function (PDF) found in a full hydrodynamic simulation. The PDFs used in this calculation are gas density PDFs extracted from two hydrodynamical simulations. Specifically, they give the fraction of gas mass in bins of overdensity \(\Delta\) at each redshift. We use these density PDFs to weight the C\,\textsc{iv} abundance predicted by our 0D model at each \(\Delta\), assuming a uniform metallicity. Integrating over the full density distribution then yields the predicted cosmic C\,\textsc{iv} density parameter, \(\Omega_{\rm CIV}\).
The density PDF makes sure that our 0D Lagrangian parcels are attributed to a relative volume in the universe at any given redshift. This allows us to expand our 0D simulations to make predictions about several quantities of the IGM, from a statistical point.

The comoving C\,\textsc{iv} density parameter for a single gas parcel with overdensity, $\Delta_j$, is
\begin{align}
    \Omega_{\textrm{CIV}}(\Delta_j, z) \equiv \frac{\rho_{\textrm{CIV}}(\Delta_j, z)}{\rho_{\textrm{crit}}} &= \frac{x_{\textrm{CIV}}(\Delta_j, z)\rho_{\textrm{C}}(\Delta_j)}{\rho_{\textrm{crit}}}\\
    &= \frac{x_{\textrm{CIV}}(\Delta_j, z)n_{\textrm{C}}(\Delta_j)m_{\textrm{C}}}{\rho_{\textrm{crit}}}\\
    &= \frac{x_{\textrm{CIV}}(\Delta_j, z)[n_{\textrm{H}}(\Delta_j)ZX_{\textrm{C}}]m_{\textrm{C}}}{\rho_{\textrm{crit}}}
\end{align}
where $X_{\textrm{C}} = 2.45\times10^{-4}$ is the abundance of carbon relative to hydrogen for a Solar metallicity gas and $Z$ is the metallicity (in solar units). Thus, the PDF-weighted prediction for the C\,\textsc{iv} mass fraction is approximated by
\begin{equation}
    \langle \Omega_{\mathrm{C\,IV}}(z)\rangle \simeq \sum_{j} w_j(z)\Omega_{\textrm{CIV}}(\Delta_j, z)
\end{equation}
By design, the approach preserves the primary density dependence with negligible expense. At the same time, it also isolates the key systematics: the chosen set $\{\Delta_j\}$, the reliability of the underlying PDF, and any evolution of $Z$ with redshift.

\begin{figure}
\centering
\includegraphics[width=\columnwidth]{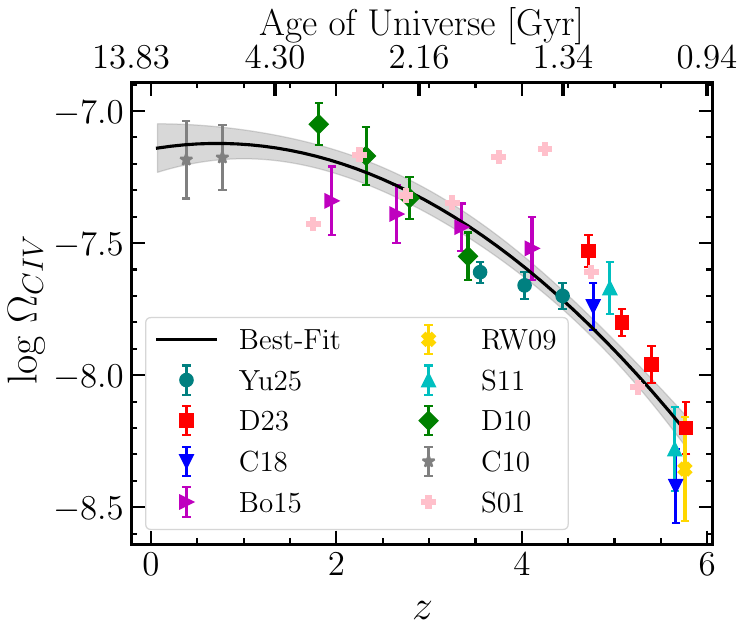}
\caption{Quadratic fit to the observed C\,\textsc{iv} density parameter, \(\log \Omega_{\rm CIV}\), as a function of redshift. Coloured symbols show individual observational constraints: \citet{Yu_2025} (teal circle), \citet{Davies_2023} (red square), \citet{Codoreanu_2018} (blue down triangle), \citet{Boksenberg_2015} (magenta right triangle), \citet{Ryan-Weber_2009} (yellow cross), \citet{Simcoe_2011} (cyan up triangle), \citet{D_Odorico_2010} (green diamond), \citet{Cooksey_2010} (gray star), and \citet{Songaila_2001} (pink plus). The black curve shows the quadratic fit, and the gray shaded region indicates the 1\(\sigma\) uncertainty.}
\label{fig:civ_fit}
\end{figure}

We compare our estimates of the C\,\textsc{iv} density parameter to an ensemble of observational measurements. For ease of comparison, we compress the observed redshift evolution of $\Omega_{\rm CIV}$ into a simple quadratic relation in the \(\{z, \log \Omega_{\rm CIV}\}\) plane (fig.~\ref{fig:civ_fit}),
\begin{equation}
    \log \Omega_{\mathrm{CIV,\,fit}}(z) = \sum_{i=0}^2 c_i z^i \, ,
\end{equation}
where the \(c_i\) are best-fitting coefficients obtained from the compilation of data. Figure \ref{fig:civ_fit} shows the fit solid line to the data points. The gray band represents the $1\sigma$ uncertainty in the fit. In future plots, we will use this fit to represent the broad behavior of the $\Omega_{\rm CIV}$ data.

\begin{figure*}
\centering
\includegraphics[width=0.8\textwidth]{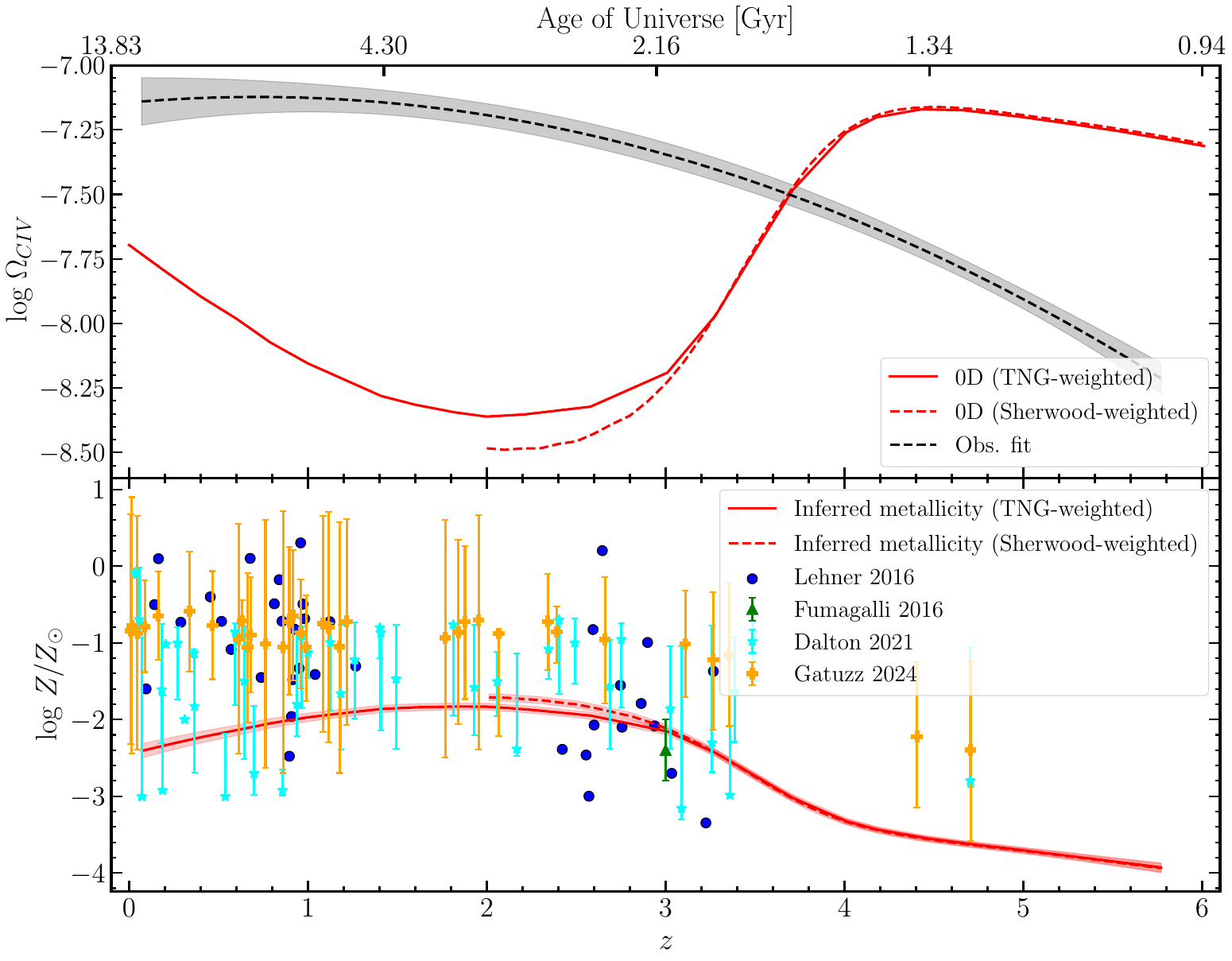}
\caption{Top: C\,\textsc{iv} density parameter, $\Omega_{\textrm{CIV}}$, obtained using our 0D framework in the IGM using full NEI thermochemistry with Illustris-TNG weights (red solid) and Sherwood weights (red dashed) assuming constant metallicity, $Z = 10^{-3} Z_{\odot}$. The black dashed curve represents the empirical fit for various observations of C\,\textsc{iv} density parameter as mentioned in fig.~\ref{fig:civ_fit}. The gray shaded region denotes the 1-$\sigma$ uncertainty. Bottom panel: Metallicities inferred for the observational samples shown in the top panel, with solid (dashed) curve showing TNG-(Sherwood-) weighted metallicities. Blue circles denote directly reported metallicities in \citet{Lehner_2016} using multiple metal lines in pLLs and LLS systems, cyan stars and orange plus markers represent IGM metallicities from blazar observations as found in \citet{Dalton_2021} and \citet{Gatuzz_2024} respectively, and the single green triangle reports the metallicity found in \citet{Fumagalli_2016}.}
\label{fig:civ_main}
\end{figure*}

The top panel of fig.~\ref{fig:civ_main} compares this empirical fit (black dashed) to the PDF-weighted cosmic C\,\textsc{iv} abundance predicted by our model, allowing a direct assessment of how well the simulated redshift evolution reproduces the observed \(\Omega_{\rm CIV}(z)\). To show that our results are insensitive to the choice of weights, we take the PDFs from two different simulations, namely, the Illustris-TNG \citep{Pillepich_2018, Nelson_2019} (red solid) and the Sherwood \citep{Bolton_2016} (red dashed) simulations.

We see that \Ociv remains high till $z\sim 4$ when the universe is moderately ionized and then declines sharply with the onset and all the way up to the end of He\,\textsc{ii} reionization.  This is because \ion{He}{ii} has a similar ionization energy as \civ, and therefore, \ion{He}{ii}$\rightarrow$ \ion{He}{iii} transition is also associated with the \civ$\rightarrow$ \ion{C}{v} transition. 
Realistically, the metals in IGM would increase with time due to outflows. This leads to a constant C\,\textsc{iv} abundance over $1 \lesssim z \lesssim 3$ reflecting a balance between outflows and He\,\textsc{ii} reionization. We suspect that the continued rise of C\,\textsc{iv} in our model after the end of He\,\textsc{ii} reionization is at least partly driven by our ignorance of additional local ionizing sources, which are likely to become increasingly important at these redshifts and overdensities. 

The mismatch in the predicted \Ociv (red lines) and the observed \Ociv (black dashed line) is mainly because of our assumption of a constant metallicity (i.e., $Z = 10^{-3} Z_\odot$) of the gas parcels. In fact, we can turn this to our advantage to estimate the amount of metals in the IGM that is required to match our predicted \Ociv to the observed one.
In the bottom panel of Figure \ref{fig:civ_main}, we translate the measurements of $\Omega_{\rm CIV}$ into metallicity inferences by comparing the observed \(\Omega_{\rm C\,IV}\) to the 0D model prediction (normalized to the fiducial \(Z/Z_{\odot}=10^{-3}\)), plotted as \(\log Z/Z_{\odot} = \log\,[\Omega_{\rm obs}/(\Omega_{\rm 0D}/10^{-3})]\). The inference using Illustris-TNG weights is shown in red solid, and the one with Sherwood is shown with red dashed. We see that the metallicity requirement in the IGM increases with time till $z\sim 2-3$ when the universe was at its peak of star formation. The metallicity requirement saturates afterwards, probably because of the declining of the star formation rate and a lack of metal escape from the galaxies as the galaxies also become more massive to trap the metals in their own circumgalactic medium (CGM). For a comparison, we also show the reported metallicity (blue circles) for partial Lyman-limit systems and Lyman-limit systems (pLLS+LLS), i.e., log $N_{\rm HI}/{\rm cm^{-2}} < 18.3$ in \citet{Lehner_2016}, the regions likely to be the primary reservoir of C\,\textsc{iv}. Cyan star and orange plus markers represent metallicity measurements from blazar observations as reported in \citet{Dalton_2021} and \citet{Gatuzz_2024}, respectively. The sole green triangle shows the metallicity reported by \citet{Fumagalli_2016}. Both inferred bands corresponding to the different surveys match the observational constraints very well at $z \sim 2-3$, while at lower redshifts, $z < 1$, we infer somewhat lower metallicities, driven by the higher C\,\textsc{iv} content discussed above.
Additionally, in this work, the abundance ratio of C\,\textsc{iii} to C\,\textsc{iv} is 1.2, which is slightly higher than the observational value of $(0.7^{+0.43}_{-0.20})$ reported in \citet{Shull_2014}. However, we overestimate the carbon ionization correction factor by approximately a factor of three. We suspect that this discrepancy may arise, at least in part, from our neglect of local galactic ionizing sources at low redshift. Nevertheless, these values remain within the substantial scatter of the observational measurements. Again, our framework remains successful in predicting the metal content in the universe.

One limitation of our framework could be the presence of shocked gas in the CGM around host galaxies. In fact, C\,\textsc{iv} resides predominantly in overdensities $\Delta \gtrsim 100$ (see section \cref{subsec:implications-to-obs}), with the maximum overdensity in our model reaching $\Delta \sim 10^3$ by $z \sim 0$. One might therefore naively identify C\,\textsc{iv} as a tracer of the circumgalactic medium. However, the C\,\textsc{iv}-bearing gas is photoionized with temperatures of a few $\times 10^4$–$10^5$~K, as indicated by their very low line width, $b_{\rm CIV}$. It is certainly possible that CGM of the host galaxies could contribute to some of the observed \Ociv. However, halos with virial temperatures in this range (typically, $M_{\rm vir} \lesssim 10^{11}$ M$_\odot$) lie in the regime where the radiative cooling timescale, $t_{\mathrm{cool}}$, is shorter than the free-fall (dynamical) timescale, $t_{\mathrm{ff}}$, i.e. $t_{\mathrm{cool}} \lesssim t_{\mathrm{ff}}$, so that a quasi-static hot atmosphere is thermally unstable and cannot be maintained \citep{Birnboim_2003}. Additionally, even if a temporary CGM is created by strong supernovae feedback in these galaxies to overcome radiative cooling, the CGM would have to be hot, and therefore, inconsistent with the observed $b_{\rm CIV}$.
Consequently, although C\,\textsc{iv} is associated with relatively dense gas in the vicinity of haloes, it is unlikely to originate mainly from a long-lived, volume-filling hot CGM phase. Instead, it appears to predominantly trace photo-ionized gas in halo outskirts, consistent with the picture supported by \citet{bromberg2026}. Our framework therefore provides a useful first-order way to infer the mean metallicity from a single metal-ion observation. However, this estimate implicitly assumes that the relevant gas at a given density is entirely photo-ionized. In reality, a fraction of the gas may instead reside in a shock-heated phase, and the resulting effect on C\,\textsc{iv} is not necessarily monotonic.
Gas heated to \(T \sim 10^5\,{\rm K}\) can enhance C\,\textsc{iv}, since at these temperatures collisions efficiently ionize lower carbon states into C\,\textsc{iv}, while the gas is not yet hot enough to ionize most carbon to higher stages such as C\,\textsc{v}; in this case, a lower metallicity could reproduce the observed \(\Omega_{\rm CIV}\). Whereas gas heated to higher temperatures, $T \sim 10^6 - 10^7\mathrm{K}$ would ionize carbon to higher stages such as C\,\textsc{v} and above, thereby reducing the C\,\textsc{iv} fraction. Thus, the inferred metallicity under a single-phase assumption may shift either upward or downward depending on the thermal distribution of the shocked gas.

\subsection{Variations in reionization \& UVB}
\label{subsec:var}
\begin{figure}
\centering
\includegraphics[width=\columnwidth]{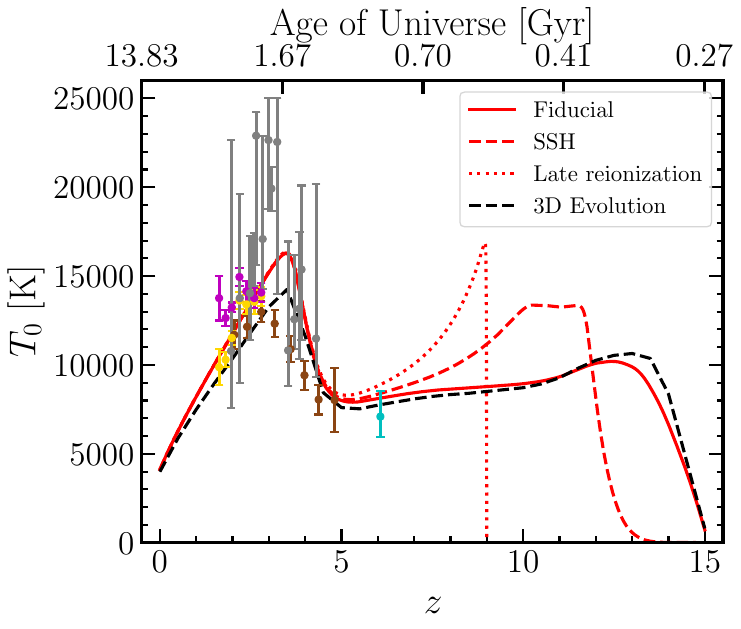}` 
\caption{Evolution of $T_0$ \textit{w.r.t.} redshift for our 0D model for fiducial and different physical cases, compared to full 3D NEI hydro simulation. Overlaid are observational constraints on $T_0$, identical to fig.~\ref{fig:T0}.}
\label{fig:T0_var}
\end{figure}

One of the hallmarks of our model is its speed and flexibility, which lets us systematically probe how key IGM ingredients shape observables. In what follows, we exploit this to vary the reionization timeline and self-shielding assumptions, which in turn affect the radiation field, and quantify their impact on the thermal and ionization histories, the cosmic C\,\textsc{iv} abundance, and the metallicities inferred from C\,\textsc{iv}. The variations are as follows:
\begin{itemize}
    \item \textbf{Reionization epoch} (Fig.~\ref{fig:T0_var}): Varying the reionization epoch is essential because the IGM retains strong \emph{thermal memory} \citep{HuiGnedin1997, Hui_2003, Tittley_2007, Nasir_2016}: once photoheating occurs, the gas cools gradually via adiabatic and radiative losses, so the timing of reionization can imprint on \(T_0(z)\), \(\gamma_{\textrm{fit}}(z)\) and, the ion fractions. Furthermore, the true nature of reionization, its onset, duration, relative contribution of ionizing sources, and spectral hardness - remains uncertain. Therefore, adopting a single history risks biasing thermal and metallicity inferences. Keeping this in mind, we try a late reionization model, where the UVB is switched on at $z \sim 9$.

    \item \textbf{Self-shielding}: At higher column densities (e.g., in LLS/DLA-like regions; $N_{\rm HI} \gtrsim 10^{18}$ \pcmsq), IGM is no longer optically thin, and the UVB begins to attenuate \citep{ME2000, Schaye_2001}. This self-shielding effect increases neutral fractions, reduces temperatures, and causes a departure from the diffuse IGM \(T\)–\(\Delta\) relation by reducing the local photoionization and photoheating rates. This, in turn, can shift metal ion balances toward lower stages (e.g., C\,\textsc{iv} over C\,\textsc{v}) and increases the contribution of dense gas to \(\Omega_{\mathrm{C\,IV}}\), thereby reducing metallicity inferences. To test the effect of self-shielding, we assume the following attenuated radiation field
    \begin{equation}
    J_\nu^{\rm eff} = J_\nu^{\rm UVB}\,e^{-\tau_{\mathrm{HI}}(\nu)},
    \label{eq:ssh}
    \end{equation}
    where \(\tau_{\mathrm{HI}}(\nu)\) is the H\,\textsc{i} optical depth given by, $\tau_{\rm HI}(\delta, \nu)=N_{\rm HI}(\delta)\sigma_{\rm HI}(\nu)$ where \(N_{\mathrm{HI}}\) is the local H\,\textsc{i} column density and \(\sigma_{\mathrm{HI}}\) is the photoionization cross-section of \ion{H}{i}. However, in our zero-dimensional runs, there are no physical sightlines and hence no well-defined absorber column densities to compute $\tau_{\rm HI}(\nu)$ from first principles. Therefore, we refer to Eq.10 in \citet{Schaye_2001}
    \begin{align}
    N_{\rm HI} &\sim\,2.7 \times 10^{13}\, \textrm{cm}^{-2}\, (1 + \delta)^{3/2}\,T_4^{-0.26}\,\Gamma_{12}^{-1}\\
    &\times\left(\frac{1+z}{4}\right)^{9/2}\left(\frac{\Omega_bh^2}{0.02}\right)^{3/2}\left(\frac{f_g}{0.16}\right)^{1/2}
    \label{eq: NHI}
    \end{align}
    to compute column density and in turn, optical depth from overdensity, temperature, and hydrogen photoionization rate ($\Gamma$). Here, $T_4 \equiv T/(10^4\, \rm K)$, $\Gamma_{12} \equiv \Gamma / (10^{-12}\,\rm s^{-1})$ is the photo-ionization rate, and $f_g$ is the baryonic fraction.
\end{itemize}

The evolution of $T_0$ with these variations is shown in Fig.~\ref{fig:T0_var} (dashed red line). In the self-shielding case, we notice a delay in ionization due to the large optical depth at $z\sim 14$ thanks to the high \ion{H}{i} fraction at the early universe. The higher \ion{H}{i} fraction initially further increases the IGM temperature after the ionization is complete. The effects of the enhanced heating persist all the way until the beginning of He\,\textsc{ii} reionization, albeit with decreasing differences. Similarly, a later reionization also causes a sharp initial rise in temperature owing to the strong redshift dependence of H\,\textsc{i} optical depth, after which temperature quickly declines to converge with the fiducial case.

We further examined the impact of self-shielding on the C\,\textsc{iv} density parameter and found it to be negligible: the resulting \(\Omega_{\mathrm{C\,IV}}\) differs from the optically thin case by less than a percent at all redshifts. This is readily understood from how self-shielding enters our calculation. The local radiation field is attenuated once the H\,\textsc{i} optical depth at the Lyman limit reaches unity, \(\tau_{\rm HI}(\nu = 13.6\,\mathrm{eV}) \sim 1\), corresponding to a column density \(N_{\rm HI} \sim 1.6\times 10^{17}\,\mathrm{cm^{-2}}\). For fiducial parameters at \(z \sim 3\), with \(T_4 \equiv T/10^4\,\mathrm{K} \sim 1\) and \(\Gamma_{12} \equiv \Gamma_{\mathrm{HI}}/10^{-12}\,\mathrm{s^{-1}} \sim 1\), Eq.~\ref{eq: NHI} reduces to
\begin{equation}
    N_{\rm HI} \simeq 2.7 \times 10^{13}\,(1+\delta)^{3/2}\,\mathrm{cm^{-2}},
\end{equation}
implying that H\,\textsc{i} self-shielding becomes important at overdensities \(\delta \sim 200\).

The ionization potential for C\,\textsc{iv} is roughly four times higher than that of H\,\textsc{i}. Since the H\,\textsc{i} photoionization cross section scales as \(\sigma_{\rm HI} \propto \nu^{-3}\), achieving \(\tau \sim 1\) at the C\,\textsc{iv} edge would require a factor of \(\sim 4^3 \simeq 64\) higher \(N_{\rm HI}\), corresponding to an overdensity larger by a factor of \(\sim 15\). Thus, C\,\textsc{iv}-bearing gas would only become self-shielded at \(\delta \sim 3000\). Gas at such high densities, however, is unlikely to host a significant fraction of C\,\textsc{iv}, since rapid recombination drives carbon into lower ionization states (also see section \ref{subsec:implications-to-obs}).

\begin{figure}
\centering
\includegraphics[width=\columnwidth]{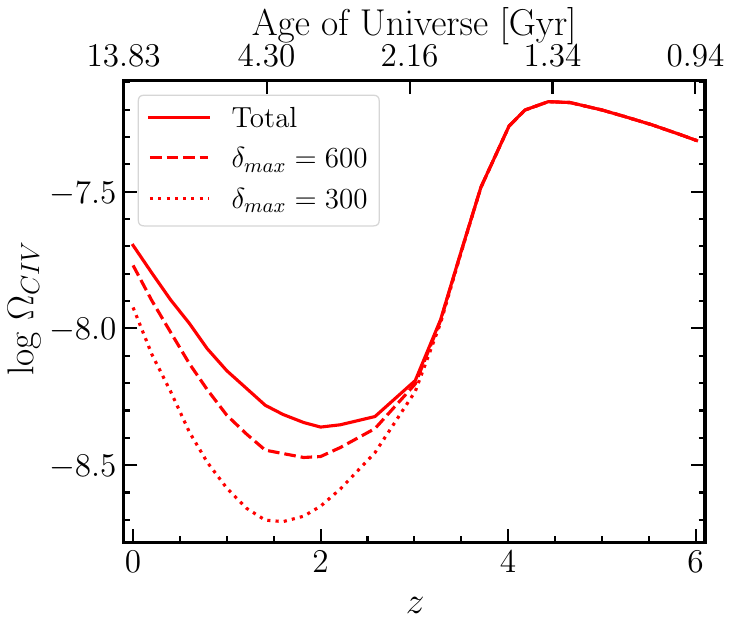}` 
\caption{C\,\textsc{iv} density parameter for three choices of maximum overdensity:
(i) the full case including gas up to \(\delta \sim 1000\) (solid),
(ii) restricted to \(\delta \lesssim 600\) (dashed), and
(iii) restricted to \(\delta \lesssim 300\) (dotted).}
\label{fig:civ_conv}
\end{figure}

We confirm this explicitly in fig.~\ref{fig:civ_conv}, where we compare the total \(\Omega_{\rm CIV}\) obtained when including gas up to a maximum density contrast \(\delta_{\rm max} \sim 1000\) at \(z \sim 0\) with two restricted cases in which we truncate the integral at \(\delta_{\rm max} = 600\) and \(\delta_{\rm max} = 300\), respectively. Limiting to \(\delta_{\rm max} \sim 300\) underestimates \(\Omega_{\rm CIV}\) by about half a dex, but extending to \(\delta_{\rm max} \sim 600\) recovers the full result to within a few per cent. It is therefore safe to conclude that essentially all of the C\,\textsc{iv} arises from gas with \(\delta \lesssim 1000\), well below the densities at which C\,\textsc{iv}-bearing gas would be significantly affected by self-shielding.

%%%%%%%%%%%%%%%%%%%%%%%%%%%%%%%%%%%%%%%%%%%
\section{Discussions}
\label{sec:discussions}

\subsection{Implications for observations} \label{subsec:implications-to-obs}

Most observational studies of C\,\textsc{iv} indicate that a large fraction of the cosmological C\,\textsc{iv} budget resides in low- to moderately overdense, photoionized gas as indicated by measurements of C\,\textsc{iv} impact parameter, $b_{\rm CIV} \sim 10 \textrm{\,km s}^{-1}$. At higher temperatures, where collisional ionization is relevant, carbon naturally shifts to higher ionization stages C\,\textsc{v}/\textsc{vi}/\textsc{vii}. These C\,\textsc{iv}-bearing regions are typically associated with the progenitors of the Ly$\alpha$ forest and Lyman-limit systems rather than the dense, virialized interiors of halos. Pixel optical depth analyses \citep{Ellison_2000, Aguirre_2002, Aguirre_2008, Schaye_2003, Turner_2016} show that C\,\textsc{iv} is routinely detected down to Ly$\alpha$-forest overdensities ($\delta \sim 10$), and that the inferred [C/H] at these densities is of order $10^{-3}$-$10^{-2}$ at $z \sim 2$-3. \citet{Kim_2013} explicitly separate ``enriched'' and ``unenriched'' \ion{H}{i} forest and find that C\,\textsc{iv}-producing gas spans a wide range of $N_{\mathrm{HI}}$, mostly in the forest / LLS regime rather than in classical damped Lyman-$\alpha$ (DLA) columns. Additionally, large blind surveys that measure the cosmic C\,\textsc{iv} density parameter $\Omega_{\mathrm{C\,IV}}$ \citep{Songaila_1997, Songaila_2001, Songaila_2005, D_Odorico_2010, D_Odorico_2013, D_Odorico_2022} mostly target intervening C\,\textsc{iv} systems identified in high-resolution QSO spectra, typically outside obvious DLAs and sub-DLAs. Yet they recover essentially the full $\Omega_{\mathrm{C\,IV}}$ quoted in the literature from $z \gtrsim 1.5$ up to the reionization epoch. This strongly indicates that DLAs and sub-DLAs do not dominate $\Omega_{\mathrm{C\,IV}}$: they host important metal reservoirs in neutral gas but are too rare in path length to supply most of the cosmological C\,\textsc{iv} mass.

\begin{figure}
\centering
\includegraphics[width=\columnwidth]{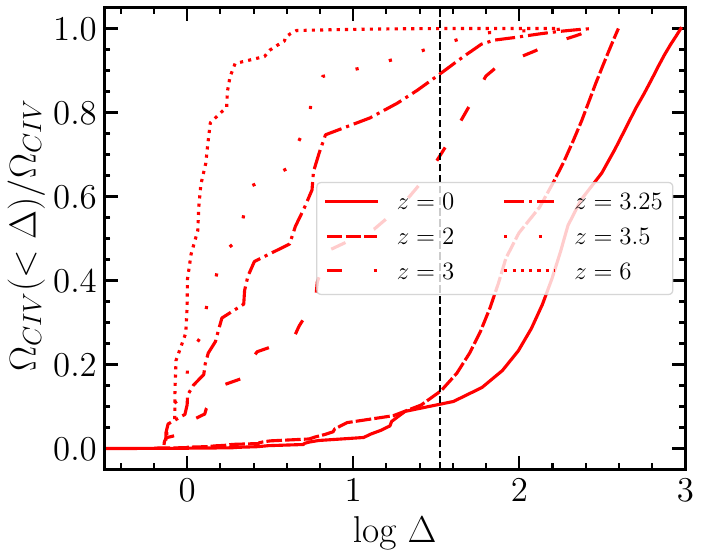}
\caption{Normalized C\,\textsc{iv} abundance (see text for definition) in our 0D model \textit{w.r.t.} overdensity at given redshifts. Black vertical line denotes \textit{virial overdensity} for a Lagrangian density point (refer to Appendix \ref{sec:appendixB} for calculation and assumptions).}
\label{fig:civ_delta}
\end{figure}

To illustrate \emph{where} in the density field the global C\,\textsc{iv} budget resides, and how this partition evolves with time in this work, we show in fig.~\ref{fig:civ_delta}, the cumulative C\,\textsc{iv} density parameter, \(\Omega_{\mathrm{C\,IV}}(<\Delta)\) normalized by total C\,\textsc{iv} density parameter, as a function of overdensity at several redshifts. As evident, the dominant source of C\,\textsc{iv} at higher redshifts is the diffuse IGM with $\Delta \lesssim 10$ regions contributing more than 50\% of the content even up to $z \sim 3$. This picture is further supported by the scarcity of massive galaxies at high redshift, which implies shallower potential wells from which metals can more easily escape, and by the smaller physical size of the Universe at early times, which facilitates the mixing of outflowing metals into the surrounding IGM \citep{Scannapieco_2002, Oppenheimer_2006, Tornatore_2007, Bromm_2011}. With time, this picture changes, as more massive halos assemble and the cosmic web becomes increasingly non-linear, an ever larger fraction of C\,\textsc{iv} originates in moderately to highly overdense environments associated with near-galaxy regions and group-scale halos rather than the diffuse IGM. By $z \sim 2$, more than half of the total C\,\textsc{iv} budget already arises from gas at $\Delta \gtrsim 100$, and by the present epoch this fraction exceeds $90\%$, indicating a progressive migration of C\,\textsc{iv} from diffuse IGM into dense, halo-linked environments.

This picture is further reinforced by the kinematics of C\,\textsc{iv} absorbers. Firstly, by construction, blind surveys identify C\,\textsc{iv} systems directly from absorption features along random quasar sightlines, without requiring any pre-selected galaxy counterpart. Galaxy associations are then established \emph{a posteriori} by searching for nearby galaxies in redshift surveys around each absorber, typically within projected separations of order a few \(\times 100\,\mathrm{kpc}\) and line-of-sight velocity offsets \(|\Delta v| \lesssim \mathrm{few}\times 100\,\mathrm{km\,s^{-1}}\) \citep[e.g.][]{Landoni_2016,Banerjee_2023}. In practice, many absorbers either have multiple plausible galaxy matches within these windows or no secure counterpart at all. Combined with the relatively high C\,\textsc{iv} covering fractions out to, and in some cases beyond, the virial radius, this strongly indicates that a substantial fraction of \(\Omega_{\mathrm{C\,IV}}\) resides in an extended, metal-enriched IGM surrounding halos, rather than being confined to the inner CGM of individual galaxies.

Secondly, if C\,\textsc{iv} predominantly traces such low-density gas, its redshift-space clustering around galaxies is expected to show only weak line-of-sight elongation on small scales, i.e., suppressed finger-of-God features in the 2D galaxy-C\,\textsc{iv} correlation function. Conversely, the clustering signal on Mpc scales may be strengthened, as C\,\textsc{iv} traces large-scale structure and filamentary accretion flows rather than the virialized motions of gas within individual haloes. In this picture, the distribution of absorber-galaxy velocity offsets, $|\Delta v|$, should also be comparatively narrow, with fewer absorbers at large $|\Delta v|$ than in a population dominated by virialized halo gas. On larger, quasi-linear scales, the redshift-space signal should be shaped mainly by coherent infall toward overdense regions rather than by random virial motions within haloes. In that case, C\,\textsc{iv} would trace the large-scale cosmic web and filamentary velocity field more closely than the disturbed internal kinematics of halo gas. Observational studies provide some suggestive support for this picture, with C\,\textsc{iv} absorbers often found within a few hundred \(\rm km\,s^{-1}\) of nearby galaxies and, in some cases, interpreted as tracing filamentary or accreting structures \citep{Banerjee_2023}. However, current evidence is not yet sufficient to establish that the redshift-space signal is generically dominated by coherent infall rather than virial motions in halo gas.

\begin{figure*}
\centering
\includegraphics[width=\textwidth]{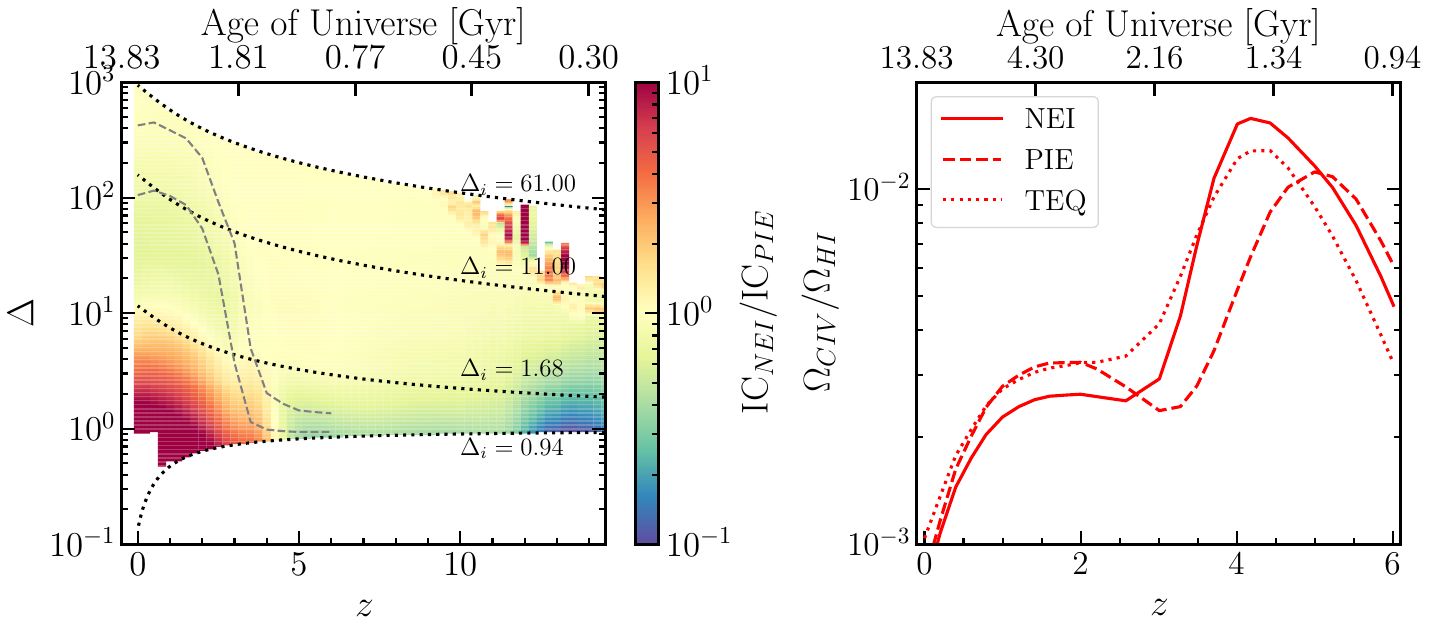}
\caption{Left: Colormap of the ratio between our ionization correction and the standard PIE-based correction used in observational metallicity estimates, \((x_{\rm CIV}/x_{\rm HI})_{\rm NEI} / (x_{\rm CIV}/x_{\rm HI})_{\rm PIE}\), shown in the \(z\)–\(\Delta\) phase space. The dotted black curves trace the trajectories of a representative set of density points used in our analysis. The gray dashed curves show overdensities containing 25\% and 75\% of global C\,\textsc{iv} content, showing why non-equilibrium does not have much impact on the overall \ion{C}{iv} density parameter for $z \lesssim 2$. The missing pixels towards low-$z$ and low-$\Delta$, and high-$z$ and high-$\Delta$ indicate extremely small ion fractions for C\,\textsc{iv} and/or H\,\textsc{i}. Right: Time evolution of global C\,\textsc{iv} density parameter relative to H\,\textsc{i} density parameter in our 0D model (NEI), under the assumption of PIE, and assuming total equilibrium (TEQ) (see text for details). The NEI treatment predicts enhanced C\,\textsc{iv} abundances relative to PIE over \(3.5 \lesssim z \lesssim 5\), broadly coincident with He\,\textsc{ii} reionization. After this phase, the C\,\textsc{iv}-bearing gas progressively relaxes toward photoionization equilibrium, and the difference between the NEI and PIE predictions decreases. In contrast, the C\,\textsc{iv} TEQ model remains in significantly better agreement with the NEI calculation both across this redshift interval and at later times.}
\label{fig:ionr}
\end{figure*}

\subsection{Ionization balance in the IGM}
\label{subsec:ionization-balance}
Given this environmental and kinematic picture, a natural next step is to ask \emph{where} in density space the usual photoionization equilibrium assumption remains valid. In particular, if most C\,\textsc{iv} arises in low- to moderately overdense, photoionized gas in the near-galaxy IGM, we need to quantify over which ranges of \(\Delta\) and redshift PIE provides an adequate description, and where non-equilibrium effects or time-dependent radiation fields become important. We therefore now turn to a detailed comparison of our 0D ionization model with standard PIE-based ionization corrections as a function of \(z\) and \(\Delta\).
Accurate modeling of the ionization correction (IC) is essential for predicting the metallicity using observations of column densities, as shown in the following equation,
\[
\left[\frac{X}{\mathrm{H}}\right]
= \log_{10}\!\left(\frac{N_{X^{i}}}{N_{\mathrm{HI}}}\right)
- \log_{10}\!\left(\frac{x_{\mathrm{X^i}}}{x_{\rm HI}}\right)
- \log_{10}\!\left(\frac{X}{\mathrm{H}}\right)_\odot ,
\]
where
\(N_{X^{i}}\) is the column density of ion \(X^{i}\) (e.g. C\,\textsc{iv}),
\(N_{\mathrm{HI}}\) is the neutral hydrogen column density,
\(x_{X^{i}} \equiv n_{X^{i}}/n_X\) and \(x_{\mathrm{HI}} \equiv n_{\mathrm{HI}}/n_{\mathrm{H}}\) are the ion fractions from a photoionization model, $(x_{X^{i}}/x_{\rm HI})$ is the ionization correction, and \((X/\mathrm{H})_\odot\) is the solar abundance, and \([X/\mathrm{H}]\) is the metallicity in dex relative to solar.

To quantify non-equilibrium effects, the left panel of Fig.~\ref{fig:ionr} shows, for each density track in our calculation, the ratio of the ionization correction predicted by the NEI model to that obtained under photoionization equilibrium,
\[
\frac{{\rm IC}_{\rm NEI}}{{\rm IC}_{\rm PIE}}
\equiv
\frac{(x_{\rm CIV}/x_{\rm HI})_{\rm NEI}}
{(x_{\rm CIV}/x_{\rm HI})_{\rm PIE}} .
\]
This ratio is evaluated along the full trajectory of each density element and then interpolated in the two-dimensional density-redshift phase space to construct a continuous colormap. To avoid regimes where the ratio is dominated by numerical noise, we restrict the plotted samples to \(x_{\rm CIV}>10^{-9}\) and \(x_{\rm HI}>10^{-7}\). With this cut, only a small number of pixels remain undefined, primarily at the high-density end and at high redshift.\\
However, the global impact of these local non-equilibrium deviations cannot be inferred directly from the left panel alone, since different density regimes contribute with different weights and may respond differently to the evolving UVB and thermal history. Since \(\Omega_x\) can be regarded, up to a normalization factor, as a density-weighted ion fraction (since it is found only after integrating over the gas density PDF), \(\Omega_{\rm CIV}/\Omega_{\rm HI}\) would roughly represent the average ion fraction ratio of \civ and \hi. The right panel of Fig.~\ref{fig:ionr} shows the redshift evolution of \(\Omega_{\rm CIV}/\Omega_{\rm HI}\) . We compare the NEI prediction with the corresponding PIE and total-equilibrium (TEQ) estimates to assess how the density-dependent ionization corrections propagate into the integrated C\,\textsc{iv}-to-H\,\textsc{i} abundance ratio. Unlike PIE, which balances only photoheating and radiative cooling, TEQ also includes the additional terms entering the full thermal evolution, namely Hubble cooling, the density-evolution term \(d\Delta/dt\), and the inverse Compton cooling. We first focus on the PIE case, since this approximation is more commonly adopted in the literature, such as \texttt{CLOUDY}-based models, and defer a more detailed discussion of the TEQ model to the next subsection.\\
The figure shows that \({\rm IC}_{\rm NEI}/{\rm IC}_{\rm PIE}\) varies with both density and redshift. Over a large part of the density-redshift phase space, the ratio remains close to unity, indicating that the NEI and PIE ionization corrections are broadly similar for much of the parameter range relevant to our calculation. The departures from unity are therefore localized to specific regimes rather than reflecting a systematic offset between the two treatments. At $z \lesssim 4$ the diffuse and under-dense gas tends to show larger ionization corrections in the NEI calculation than in PIE, corresponding to higher \(x_{\rm CIV}/x_{\rm HI}\). This behaviour arises because recombination timescales in under-dense regions are relatively long compared to the timescales over which the UVB and thermal states evolve, so the gas does not fully relax to instantaneous photoionization equilibrium and can remain over-ionized relative to the PIE solution. At higher densities, recombination becomes more efficient, allowing the ion fractions to more closely track the instantaneous ionization assumption in PIE. In these regions, the NEI and PIE ionization corrections either converge or, in some parts of the phase space, the NEI correction becomes smaller than the PIE value. However, this behaviour should not be interpreted solely as evidence that long ionization or recombination timescales drive the gas far from equilibrium. For the C\,\textsc{iv}-bearing gas considered here, the relevant photoionization and recombination timescales are generally shorter than both the timescale over which the UVB evolves and the Hubble time. Thus, the gas can remain close to an ionization-equilibrium state even when the NEI and PIE predictions differ.\\
This point is particularly relevant around \(3.5 \lesssim z \lesssim 5\), where the NEI and PIE predictions show localized departures in the ionization-correction map. At \(z\simeq 5\), the NEI and PIE carbon ion fractions are broadly similar, with most of the carbon residing in C\,\textsc{iii}. At lower redshifts, as the UVB and thermal state evolve through the He\,\textsc{ii} reionization era, the thermal history followed by the NEI calculation begins to differ from that assumed in the PIE calculation mainly due to the Hubble cooling term present in the NEI calculation. This changes the relative populations of C\,\textsc{iii}, C\,\textsc{iv}, and higher ionization states such as C\,\textsc{v}, C\,\textsc{vi}, and C\,\textsc{vii}, as shown in Fig.~\ref{fig:ion_frac_C}. The resulting enhancement or suppression of C\,\textsc{iv} in the NEI case therefore reflects both the thermal history of the gas and the ionization balance associated with that thermal state.\\
This behaviour helps explain the structure seen in the ionization-correction ratio. Regions with \({\rm IC}_{\rm NEI}/{\rm IC}_{\rm PIE}>1\), including parts of the \(3.5 \lesssim z \lesssim 5\) regime, occur where the NEI calculation yields an enhanced C\,\textsc{iv} abundance relative to PIE. Conversely, regions with \({\rm IC}_{\rm NEI}/{\rm IC}_{\rm PIE}<1\), including parts of the high-redshift phase space, indicate that the NEI ionization correction is smaller than the PIE value. These regions can respectively raise or lower the predicted contribution of C\,\textsc{iv} relative to the PIE expectation.\\
We caution, however, that the ionization-correction ratio alone should not be over-interpreted. In particular, \(\Omega_{\rm CIV}/\Omega_{\rm HI}\) depends on both the carbon and hydrogen ionization states. A lower value of \(\Omega_{\rm CIV}/\Omega_{\rm HI}\) in NEI may arise from a suppression of C\,\textsc{iv}, an enhancement of H\,\textsc{i}, or a combination of both relative to PIE. Similarly, uncertainties in the weighting and interpolation of the ion fractions over the density field and redshift can affect the precise magnitude of the difference. In the next subsection, we compare NEI, PIE, and TEQ more directly in order to separate genuine time-dependent ionization effects from differences caused by the assumed thermal state.

\subsection{Non-equilibrium ionization vs photo-ionization equilibrium vs total equilibrium}
\label{sec:teq_comparison}

\begin{figure}
\centering
\includegraphics[width=\columnwidth]{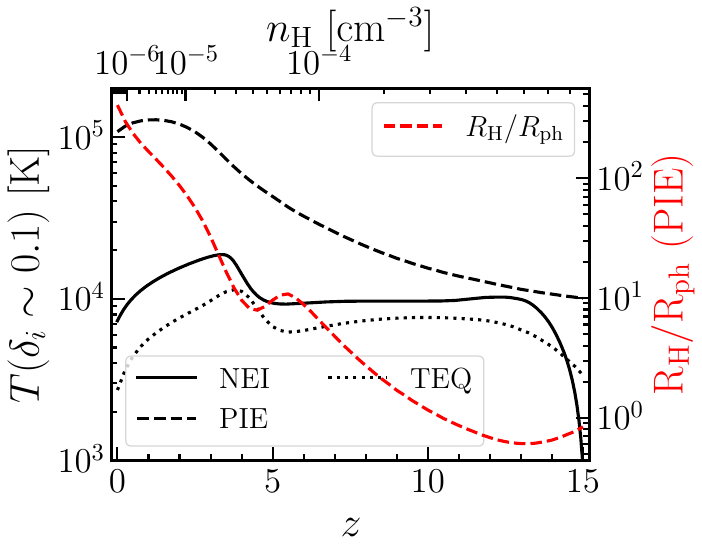}
\caption{Left axis: Black curves show the temperature evolution along a single density trajectory with initial overdensity \(\Delta_i \simeq 1.1\), corresponding to \(\Delta(z=3.5)\simeq 1.5\), for the NEI (solid), PIE (dashed), and TEQ (dotted) cases. Right axis: The red dashed line shows the ratio of the Hubble cooling rate ($R_{\rm H}$; erg s$^{-1}$ cm$^{-3}$) to the photoheating rate ($R_{\rm ph}$; erg s$^{-1}$ cm$^{-3}$) under PIE. One needs a higher photo-heating rate (by lowering the equilibrium temperature to have a higher \hi density), to maintain an ionization equilibrium at lower redshifts.}
\label{fig:rates_eq}
\end{figure}

This comparison is useful because the differences between NEI and PIE are not necessarily caused only by delayed ionization or recombination. They can also arise because the two calculations correspond to different thermal states.\\
To illustrate this, we consider gas with overdensity \(\Delta \sim 10\) at \(z\sim 3.5\), a regime that contributes significantly to C\,\textsc{iv} absorption at these redshifts. The relevant timescales are
\[
t_{\rm ph} \equiv \Gamma_{\rm CIV}^{-1},
\;
t_{\rm rr} \equiv \left(\alpha_{\rm CV}\,n_e\right)^{-1},
\;
t_{\rm rad} \equiv \left|\frac{d\ln \Gamma_{\rm CIV}}{dt}\right|^{-1},
\;
t_{\rm H} \equiv H^{-1},
\]
where \(t_{\rm ph}\) and \(t_{\rm rr}\) are the photoionization and recombination timescales of C\,\textsc{iv} and C\,\textsc{v}, respectively, \(t_{\rm rad}\) is the characteristic UVB-evolution timescale, and \(t_{\rm H}\) is the Hubble time. For the C\,\textsc{iv}-bearing gas considered here, we estimate
\[
t_{\rm ph} \sim 10^{7}\,{\rm yr}, \;
t_{\rm rr} \sim 3\times10^{6}\,{\rm yr}, \;
t_{\rm rad} \sim 5\times10^{8}\,{\rm yr}, \;
t_{\rm H} \sim 10^{9}\,{\rm yr},
\]
so that
\[
t_{\rm rr} \lesssim t_{\rm ph} \ll t_{\rm rad} \lesssim t_{\rm H}.
\]
This hierarchy shows that the ionization state can adjust much faster than either the UVB or the cosmological background evolves. Thus, the gas can remain close to ionization equilibrium.\\
The difference between PIE and TEQ is instead mainly thermal. PIE obtains the equilibrium temperature by balancing photoheating against radiative cooling alone. TEQ, on the other hand, also includes the additional terms entering the full thermal evolution, including Hubble cooling, the density-evolution term \(d\Delta/dt\), and inverse Compton cooling. Towards lower redshift, Hubble cooling becomes increasingly important relative to photoheating, as shown in Fig.~\ref{fig:rates_eq}. Here \(R\) denotes a volumetric cooling/heating rate (erg s$^{-1}$ cm$^{-3}$), i.e. a contribution to the change in thermal energy per unit volume per unit time and should not be interpreted as the inverse of a microscopic timescale. The Hubble cooling term is
\[
R_{\rm H} = 3\gamma H P ,
\]
where \(H\) is the Hubble parameter and \(P\) is the gas pressure. The photoheating contribution is
\[
R_{\rm ph} = (\gamma-1)\sum_i n_i \mathcal{H}_i ,
\]
where the sum runs over all the ions considered (
\(i=\mathrm{H\,\textsc{i}},\mathrm{He\,\textsc{i}},\mathrm{He\,\textsc{ii}}\), etc), \(n_i\) is the number density of species \(i\), and \(\mathcal{H}_i\) is the photoheating rate per ion. Thus, although the microscopic photoionization timescale can be much shorter than the Hubble time, the total volumetric photoheating rate can still be small when the relevant absorbing species are rare, as is the case once hydrogen and helium are highly ionized.

To maintain thermal equilibrium in the presence of this additional cooling channel, the gas settles to a lower temperature, producing a higher \ion{H}{i} fraction, which further contributes to higher photo-ionization heating. 
Thus, both PIE and TEQ describe equilibrium ionization states, but at different equilibrium temperatures. The closer agreement of TEQ with NEI around \(z\sim 3.5\) therefore suggests that much of the PIE-NEI discrepancy is driven by the thermal state assumed in the equilibrium calculation, rather than by a breakdown of ionization equilibrium. It also suggests that the total equilibrium is an acceptable equilibrium in the IGM ionization calculations.

\subsection{Effect of patchy reionization}
\label{subsec:patchy-eor}
Throughout this work, we have assumed a homogeneous UVB. However, observations have shown that hydrogen did not reionize everywhere at once \citep{Furlanetto_2004, McQuinn_2015, Becker_2015, Mesinger_2016}. Ionization fronts from galaxies and quasars carved expanding bubbles that overlapped over a range of times, so each region has its own reionization redshift and spectral hardness history. This spatial-temporal inhomogeneity is crucial for modeling the IGM thermal history because the impulsive heating at reionization (to $\sim 1$-$3\times 10^{4}\,\mathrm{K}$) and the long cooling time imprint a \emph{thermal memory}, increasing $T-\Delta$ scatter and driving the gas towards an isothermal state and altering pressure-smoothing scales \citep{HuiGnedin1997, Schaye_2000, Hui_2003, Gleser_2005, McQuinn_2009, Kulkarni_2015, Nasir_2016, Rorai_2017}. The same patchiness also shapes inferred metallicities and the cosmic C\,\textsc{iv} abundance: temperature variations directly change recombination rates, fluctuations in UV background hardness shift the carbon charge balance (C\,\textsc{ii}/C\,\textsc{iii}/C\,\textsc{iv}), and neutral/self-shielded islands suppress C\,\textsc{iv} even at fixed metal mass. Ignoring patchiness, therefore, biases metallicity estimates (low or high depending on local conditions) and can misestimate $\Omega_{\mathrm{C\,IV}}$ and ion ratio statistics. Hence, realistic models must track reionization timing and spatial UVB variations to connect observations of C\,\textsc{iv} to the true metal content of the IGM. This is significantly complicated to implement in our 0-D framework since a relation between local overdensities and the mass of the host galaxies has to be established first, a job that we keep for the near future.

\subsection{Local galactic radiation field}
\label{subsec:loc-rad--field}
A strong local galactic radiation field (from a nearby star-forming galaxy or QSO) raises both the photoheating and the photoionization rates, thereby reshaping the thermal and ionization histories of the surrounding CGM/IGM gas \citep{Cantalupo_2012, Oppenheimer_2013, Tumlinson_2017, Eilers_2017, Davies_2019}. The extra photoheating drives gas to higher temperatures immediately after illumination and delays subsequent cooling, increasing the pressure-smoothing scale. When the source fades, finite recombination/cooling times leave non-equilibrium (“fossil”) zones that remain over-ionized for Myr.
The extra ionizing photons from stars and AGNs can efficiently drive C\,\textsc{iii}\(\rightarrow\)C\,\textsc{iv} and, if very hard/hot, can push C\,\textsc{iv}\(\rightarrow\)C\,\textsc{v}. Consequently, the net impact on C\,\textsc{iv} is environment-dependent: moderate heating with modest spectral hardness can \emph{increase} \(x_{\rm C\,IV}\) and the C\,\textsc{iv} incidence (or the covering fraction), while very hot gas or extremely hard spectra can \emph{deplete} C\,\textsc{iv} in favor of higher ions. For metallicity inferences, a local field thereby alters the ionization corrections. Erroneous estimates of C\,\textsc{iv} under the assumption of a spatially uniform UVB may overestimate (or underestimate) metallicity depending upon spectral hardness. 
In fact, one can turn our 0-D framework into a tool to put constraints on the local radiation field for a region with known metallicity. For example, assuming that the large-scale \civ observations (as shown in Figure \ref{fig:civ_fit}) and its associate $Z(z)$ model (red lines in figure \ref{fig:civ_main}) represent the true nature of the IGM metals and the only difference in the local universe ($z\sim 0$, as measure by \cite{Lehner_2016}) is the radiation field. In that case, one needs $\sim 10$ times more local radiation field at $E>64$ eV. In the real world, of course, one would want to fit these two quantities (metallicity and the radiation field) simultaneously to the observed data.

%%%%%%%%%%%%%%%%%%%%%%%%%%%%%%%%%%%%%%%%%%%%%%%
\section{Conclusions}
\label{sec:conclusion}

We have developed and validated a metals-inclusive, non-equilibrium, zero-dimensional (0D) framework that evolves the coupled thermal and ionization state of Lagrangian gas elements exposed to a time-dependent metagalactic UV background. The model integrates stiff rate equations for H/He and a network of metal ions, while self-consistently accounting for photo-heating and other dominant cooling processes. Its central advantage is that it retains the \emph{history dependence} of ionization balance and thermal evolution (``memory'') at negligible computational cost relative to full 3D NEI hydrodynamical simulations, enabling rapid exploration of UVB choices, reionization timelines, shielding prescriptions, and density histories.

When benchmarked against full 3D NEI hydrodynamical calculations, the 0D model reproduces the IGM thermal and ionization histories with good fidelity. The mean-density temperature $T_0(z)$ is recovered to within a few $\times 10^3\,\mathrm{K}$ (up to $\sim 40\%$) over a wide redshift range, including the characteristic heating feature associated with He\,\textsc{ii} reionization. The slope of the temperature--density relation, $\gamma_{\rm fit}(z)$, is reproduced to within a few $\sim 10\%$ for $\Delta \lesssim 10$, with most residual disagreements concentrated at $z \gtrsim 6$. The redshift evolution of the H\,\textsc{i} and He\,\textsc{ii} fractions likewise tracks the 3D NEI results closely, showing that the simplified 0D treatment captures the dominant ionization-time-scale effects relevant for Ly$\alpha$-forest applications.

By evolving an ensemble of overdensities $\{\Delta_j\}$ and weighting them with baryon density probability distribution functions calibrated from cosmological hydrodynamical simulations, we obtain efficient predictions for the cosmic C\,\textsc{iv} density parameter, $\Omega_{\rm CIV}(z)$, and use them to infer metallicities from observed $\Omega_{\rm CIV}$. The predicted $\Omega_{\rm CIV}$ is dominated by diffuse-to-moderately overdense IGM gas rather than by a predominantly CGM phase. The density range that contributes to the $N_{\rm CIV}$ most evolves with time, shifting from low-density regions at high redshift to progressively more moderately overdense environments toward lower redshift. This trend is quantified by controlled density-cut experiments: truncating the density support at $\delta_{\max}\!\sim\!300$ underestimates $\Omega_{\rm CIV}$ by $\simeq 0.5$ dex, whereas extending to $\delta_{\max}\!\sim\!600$ recovers the full result to within a few percent, with substantially higher overdensities contributing little to the global C\,\textsc{iv} budget. The metallicities inferred by matching $\Omega_{\rm CIV}^{\rm obs}$ are in reasonable agreement with observational constraints (from multiple metal lines) at $z\simeq 2$-3; at low redshift, we infer somewhat lower metallicities that nevertheless remain consistent within the observational scatter.

The same framework enables rapid, transparent sensitivity tests that map astrophysical assumptions into both thermal and metal-line observables. In particular, varying the timing of reionization produces transient differences in the IGM thermal state, but both $T_0(z)$ and $\gamma_{\rm fit}(z)$ converge to the fiducial evolution by $z \sim 5$, indicating little residual ``memory'' at later times in these global diagnostics. We also find that including self-shielding has no discernible impact on the global $\Omega_{\rm CIV}$, since the highest-density, self-shielded gas contributes negligibly to the cosmic C\,\textsc{iv} mass budget in our calculations.

The present approach adopts a number of simplifying assumptions, most notably a spatially uniform UV background and prescribed density histories in place of fully coupled hydrodynamics, in exchange for computational speed and physical transparency. Its principal strength is that it provides a fast and flexible framework for following the complete ionization state of IGM gas, including H/He and an extended metal-ion network, self-consistently with the thermal evolution. Such ion-by-ion, time-dependent modeling is difficult to carry out routinely in full 3D cosmological simulations, making the present framework especially useful for exploring the impact of microphysical assumptions in a controlled manner.
Although in this work we have focused on C\,\textsc{iv} because of the abundance of observational datasets across a wide range of redshifts, the framework is sufficiently general to be applied to a broader range of metal absorbers in the IGM. A natural next application is Si\,\textsc{iv}, which traces broadly similar photoionized gas while providing complementary sensitivity to the ionization state, metallicity, and spectral shape of the ultraviolet background \citep{Louise_2025}. It would also be valuable to extend the model to ions such as O\,\textsc{vi}, which can trace a broader and physically distinct range of environments, including both photoionized gas and warmer, lower-density material in which collisional ionization may become important, but is not yet included in the present implementation. Even though different absorbers can trace environments distinct from those emphasized here, the flexibility of our framework makes it well-suited for incorporating the relevant additional physics.
More generally, applying the model jointly to multiple ionic species would enable a more complete characterization of the thermal, ionization, and enrichment state of diffuse cosmic gas. In this sense, the framework serves as a compact test bed for isolating how specific astrophysical assumptions and pieces of microphysics propagate into ion fractions and thermal histories. Natural extensions, which we defer to future work, include the incorporation of additional physics and sources of inhomogeneity such as dust heating, local galactic radiation fields, patchy reionization, and UVB fluctuations, as well as redshift-dependent enrichment histories. These developments would further strengthen the connection between microphysical modeling and absorption-line inferences.

\section*{Data Availability}

The code and data underlying this work will be shared on reasonable request to the authors.

\section*{Acknowledgment}

We would like to sincerely thank Prof. Sowgat Muzahid, IUCAA Pune, Prof. Tirthankar Roy Choudhury, NCRA-TIFR Pune, and Prof. Biman Nath, for their helpful discussions and insights during the course of this work. Their thoughtful comments, suggestions, and willingness to share ideas were very valuable and greatly improved our understanding of several aspects of the project. 
We also thank the anonymous referee for their insightful comments that improved the quality of the paper.
%%%%%%%%%%%%%%%%%%%% REFERENCES %%%%%%%%%%%%%%%%%%

\bibliographystyle{mnras}
\bibliography{references}

%%%%%%%%%%%%%%%%%%%%%%%%%%%%%%%%%%%%%%%%%%%%%%%%%%

%%%%%%%%%%%%%%%%% APPENDICES %%%%%%%%%%%%%%%%%%%%%

\appendix

\section{Comparison of IGM processes}

\begin{figure*}
\centering
\includegraphics[width=0.8\textwidth]{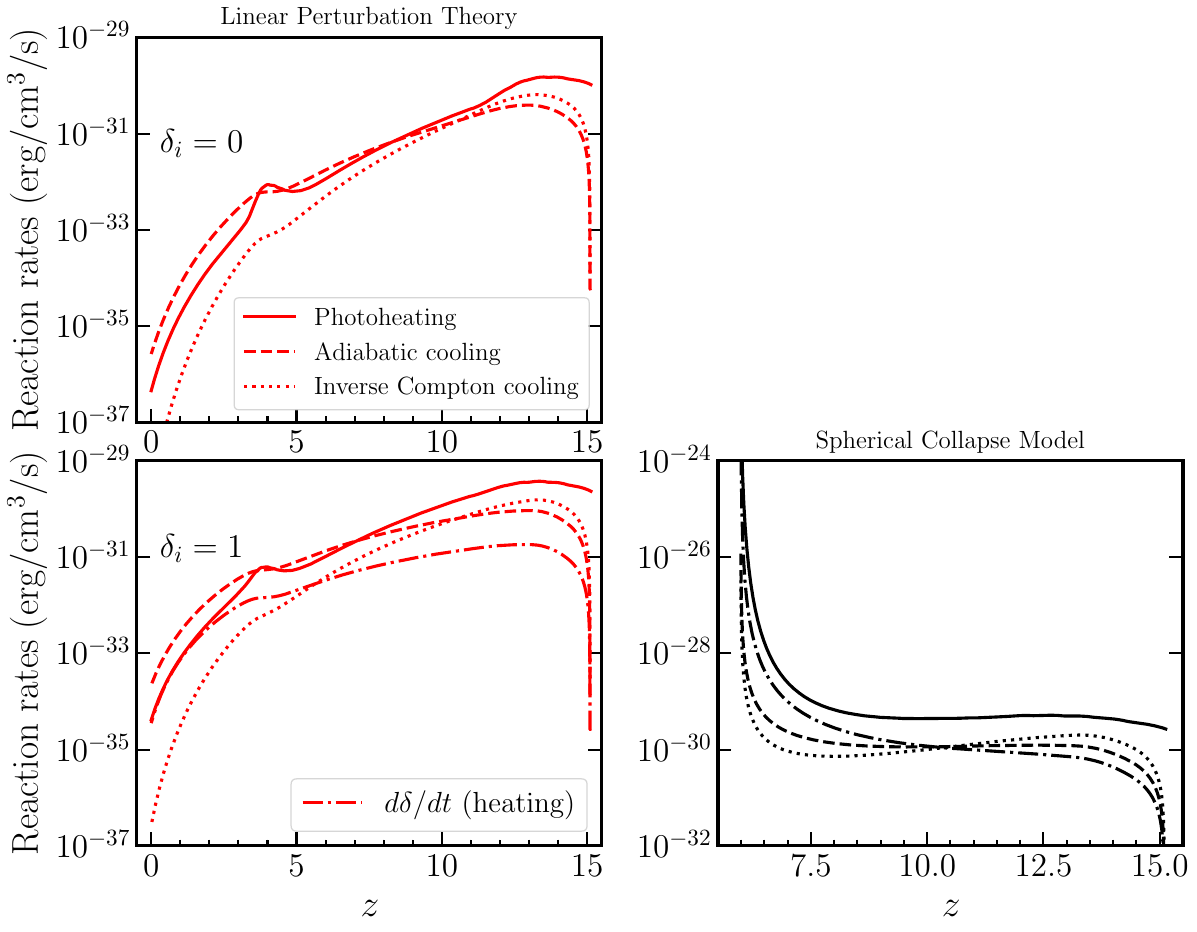}
\caption{Comparison of the heating and cooling rates employed in this study for (i) gas at the cosmic mean density (top panel) and (ii) at initial overdensity, $\delta_i=1$ (bottom panel). The left column shows density trajectory evolved using 1LPT while in the right column density grows using SCM. We stop the spherical collapse prescription at $z \sim 6$ due to shell crossing.}
\label{fig:rr}
\end{figure*}

Fig.~\ref{fig:rr} shows the evolution of the main IGM heating and cooling rates for (i) a homogeneous universe with $\delta_i = 0$ (top panel) and (ii) an initially overdense region with $\delta_i = 1$, evolved using 1LPT (bottom left) and the spherical collapse model (bottom right). At the cosmic mean density, photoheating dominates over both adiabatic and inverse Compton cooling during H\,\textsc{i}+He\,\textsc{i} reionization. Thereafter, the gas settles into photoionization equilibrium until the onset of He\,\textsc{ii} reionization. Once helium becomes predominantly doubly ionized, the thermal evolution is governed mainly by adiabatic cooling associated with the Hubble expansion.

For the overdense region, the 1LPT case shows a broadly similar qualitative evolution, with only modest differences from the mean-density case. The main distinction is the appearance of adiabatic heating from local gravitational compression, which becomes comparable to photoheating after the end of He\,\textsc{ii} reionization ($z \lesssim 3$). The spherical collapse case, however, exhibits a qualitatively different late-time behavior. At early times, when the overdensity remains in the quasi-linear regime, the evolution resembles that in 1LPT. As the collapse proceeds, however, the density grows increasingly rapidly until shell crossing is reached. In this case, photoheating remains the dominant process throughout the evolution.

\begin{figure}
\centering
\includegraphics[width=\columnwidth]{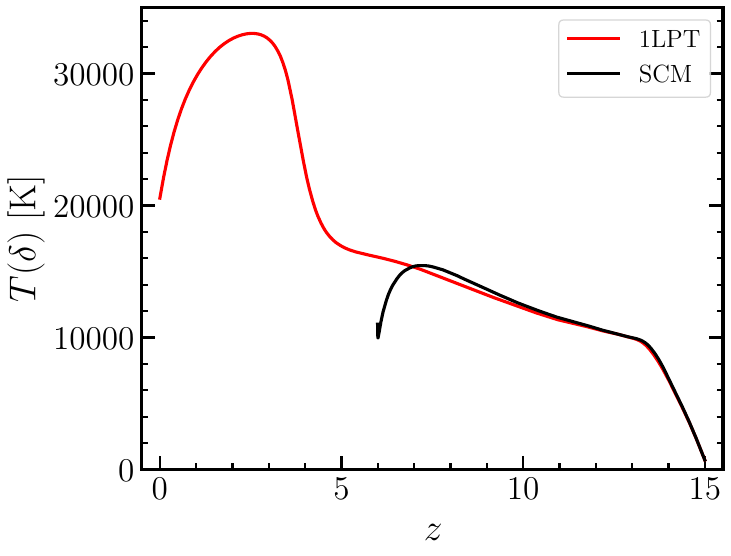}
\caption{Evolution of temperature at initial overdensity, $\delta_i=1$ using 1LPT and SCM.}
\label{fig:T_delta}
\end{figure}

Fig.~\ref{fig:T_delta} shows the corresponding temperature evolution for the $\delta_i=1$ trajectory in the 1LPT and SCM prescriptions. The two temperature tracks initially agree closely, but the spherical collapse solution drops sharply around $z \sim 7$. We speculate that this decline is caused by enhanced radiative recombination cooling, which is not shown in Fig.~\ref{fig:rr}. Because recombination cooling scales approximately as $\Delta^2$, whereas photoheating scales only as $\Delta$, it can become dominant once the overdensity grows sufficiently large, thereby driving the temperature downward.

\section{Local virialized density}
\label{sec:appendixB}

In this section, we present the calculations for local virial overdensity as shown in Fig.~\ref{fig:civ_delta}. We define the virial radius of a halo at redshift $z$ by requiring that the mean density inside $R_{\rm vir}$ is a fixed overdensity $\Delta_{\rm vir}$ times some reference density $\rho_{\rm ref}(z)$, usually either the critical density $\rho_{\rm crit}(z)$ or the mean background matter density $\bar\rho_{\rm m}(z)$:
\begin{equation}
  \bar\rho(<R_{\rm vir}) \equiv
  \frac{3 M_{\rm vir}}{4\pi R_{\rm vir}^3}
  = \Delta_{\rm vir}\,\rho_{\rm ref}(z).
\end{equation}
For a given halo mass $M_{\rm vir}$ and choice of $\Delta_{\rm vir}$, this immediately gives
\begin{equation}
  R_{\rm vir}
  = \left(
      \frac{3 M_{\rm vir}}{4\pi \Delta_{\rm vir}\,\rho_{\rm ref}(z)}
    \right)^{1/3}.
\end{equation}
We refer to $\bar\rho_{\rm vir} \equiv \bar\rho(<R_{\rm vir})$ as the \emph{virial density}.

\subsection{NFW profile and the virial density}

For an NFW halo \citep{Navarro_1997}, the density profile is
\begin{equation}
  \rho(r) = \frac{\rho_{\rm s}}{x(1+x)^2}, \qquad
  x \equiv \frac{r}{r_{\rm s}},
\end{equation}
where $r_{\rm s}$ is the scale radius and $\rho_{\rm s}$ the characteristic density. Introducing the concentration parameter $c \equiv R_{\rm vir}/r_{\rm s}$, the enclosed mass is
\begin{equation}
  M(<r) =
  4\pi \rho_{\rm s} r_{\rm s}^3
  \left[ \ln(1+x) - \frac{x}{1+x} \right].
\end{equation}
The mean density inside radius $r$ then becomes
\begin{equation}
  \bar\rho(<r)
  = \frac{3 M(<r)}{4\pi r^3}
  = 3 \rho_{\rm s}\,
    \frac{\ln(1+x) - x/(1+x)}{x^3}.
\end{equation}
Evaluating at $r = R_{\rm vir}$ (i.e.\ $x=c$) and imposing the virial definition $\bar\rho(<R_{\rm vir}) = \Delta_{\rm vir}\,\rho_{\rm ref}(z)$ gives
\begin{equation}
  \Delta_{\rm vir}\,\rho_{\rm ref}(z) =
  3 \rho_{\rm s}\,
  \frac{\ln(1+c) - c/(1+c)}{c^3},
\end{equation}
so that the scale density is
\begin{equation}
  \rho_{\rm s}
  = \frac{\Delta_{\rm vir}\,\rho_{\rm ref}(z)}{3}\,
    \frac{c^3}{\ln(1+c) - c/(1+c)}.
  \label{eq:rhos_from_Delta_vir}
\end{equation}
Equations~(\ref{eq:rhos_from_Delta_vir}) and the definition of $R_{\rm vir}$ fully specify the NFW parameters $(\rho_{\rm s}, r_{\rm s})$ for a halo of given $(M_{\rm vir}, \Delta_{\rm vir}, \rho_{\rm ref}, c)$.

\subsection{Local versus average virial overdensity}

For a given radius $r$ inside the halo, we distinguish:

\begin{itemize}
  \item the \emph{local} overdensity
  \begin{equation}
    \Delta_{\rm loc}(r) \equiv
    \frac{\rho(r)}{\rho_{\rm ref}(z)} ,
  \end{equation}

  \item the \emph{mean} overdensity enclosed within $r$
  \begin{equation}
    \bar\Delta(r) \equiv
    \frac{\bar\rho(<r)}{\rho_{\rm ref}(z)} .
  \end{equation}
\end{itemize}

For the NFW profile, the ratio of local to mean density at radius $r$ is
\begin{equation}
  \frac{\rho(r)}{\bar\rho(<r)}
  = \frac{ x^2 }
         { 3 (1+x)^2
           \left[\ln(1+x) - x/(1+x)\right] },
  \qquad x = \frac{r}{r_{\rm s}}.
  \label{eq:rho_over_rhobar_general}
\end{equation}
Evaluating equation~(\ref{eq:rho_over_rhobar_general}) at the virial radius, $x = c$, and using $\bar\Delta(R_{\rm vir}) = \Delta_{\rm vir}$, we obtain the \emph{local virial overdensity}:
\begin{equation}
  \Delta_{\rm loc}(R_{\rm vir})
  = \frac{\rho(R_{\rm vir})}{\rho_{\rm ref}(z)}
  = \Delta_{\rm vir}\,
    \frac{c^2}
         { 3 (1+c)^2
           \left[\ln(1+c) - c/(1+c)\right] }.
  \label{eq:Deltaloc_Rvir}
\end{equation}
Equivalently,
\begin{equation}
  \frac{\Delta_{\rm loc}(R_{\rm vir})}{\Delta_{\rm vir}}
  = \frac{\rho(R_{\rm vir})}{\bar\rho(<R_{\rm vir})}
  = \frac{c^2}
         { 3 (1+c)^2
           \left[\ln(1+c) - c/(1+c)\right] }.
\end{equation}

For a Milky Way halo with typical concentration, $c \sim 10$, this ratio is $\rho(R_{\rm vir})/\bar\rho(<R_{\rm vir}) \sim 0.2$, so that the local density at the virial boundary is substantially \emph{lower} than the average density inside $R_{\rm vir}$.  In other words, even though the \emph{average} virial overdensity is $\Delta_{\rm vir}\simeq 180$\footnote{The virial overdensity is known to be weakly dependent on cosmology and redshift. Given that we are only interested in simple and qualitative comparison, we assume $\Omega_{\textrm{m}}=1$ and $z=2$ for calculating $\Delta_{\textrm{loc}}(R_{\textrm{vir}})$.}, the \emph{local} overdensity experienced by a Lagrangian mass shell that ends up near $R_{\rm vir}$ is only of order $\sim 30$-40 for an NFW halo with realistic $c$.

\section{Ionization states of carbon in NEI and PIE}

\begin{figure*}
\centering
\includegraphics[width=\textwidth]{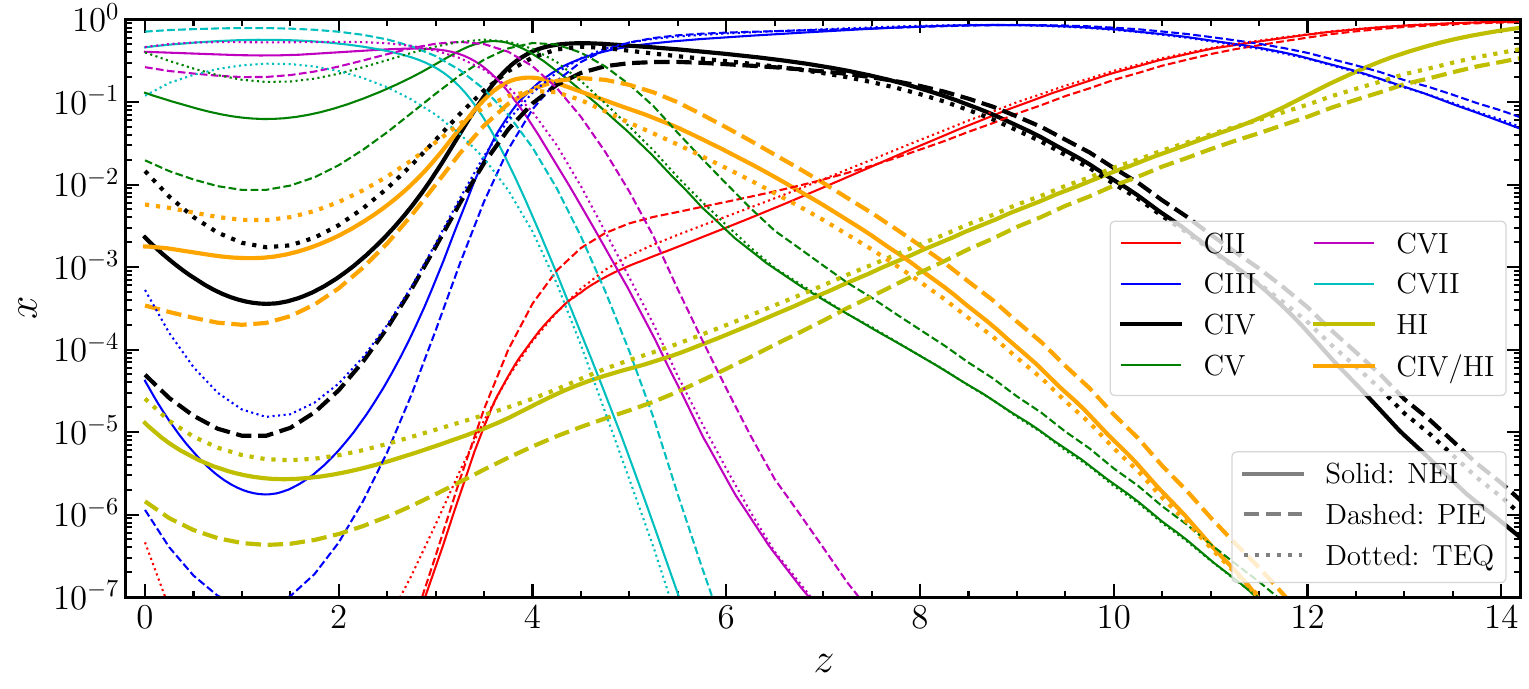}
\caption{Ionization states of carbon corresponding to initial over-density, $\Delta_i \approx 1.1$ ($\Delta \approx 1.5$ at $z\approx 3.5$), one of the initial densities likely to host the majority of C\,\textsc{iv} at $z \sim 3.5$. Additionally, we also show H\,\textsc{i} (olive green) and the ratio of ion fractions of C\,\textsc{iv} and H\,\textsc{i} normalized by $10^5$ (orange). Solid, dashed, and dotted curves denote NEI, PIE, and TEQ cases, respectively.}
\label{fig:ion_frac_C}
\end{figure*}

In Fig.~\ref{fig:ion_frac_C}, we show the evolution of different ionization states of carbon under both NEI (solid), PIE (dashed), and TEQ (dotted) for an initial overdensity of \(\Delta_i \approx 1.1\). We omit C\,\textsc{i}, as its contribution is negligible. This density is chosen because gas in this regime is expected to contribute a substantial fraction of C\,\textsc{iv} over \(z \sim 3.5\). The figure shows that at high redshifts, \(z \gtrsim 4\), all the three calculations predict C\,\textsc{iii}/\textsc{iv} to be the dominant carbon ionization states. Around the epoch of He\,\textsc{ii} reionization, however, the treatments begin to show differences. In the PIE case, the carbon ion fractions respond instantaneously to the evolving thermal and ionizing background, shifting a larger fraction of carbon into higher ionization states, especially C\,\textsc{v}, C\,\textsc{vi}, and C\,\textsc{vii}. In contrast, the NEI calculation retains memory of the previous ionization state, so the transition to these higher ions is delayed. As a result, NEI predicts relatively larger fractions of C\,\textsc{iii} and C\,\textsc{iv} over \(z \gtrsim 4\) compared to PIE. At later times, once the ionization and recombination timescales become short enough relative to the evolution of the UVB and thermal state, the NEI ion fractions relax toward their corresponding equilibrium values.
We also show a total-equilibrium (TEQ) case, in which the equilibrium temperature is computed by balancing not only photoheating and radiative cooling, but also the cosmological terms that enter the thermal evolution, including Hubble cooling, the density-evolution term \(d\Delta/dt\), and inverse Compton cooling. Including these additional thermal terms shifts the equilibrium temperature closer to the NEI thermal state, and therefore brings the corresponding carbon ion fractions into better agreement with the NEI calculation than in the PIE case.

%%%%%%%%%%%%%%%%%%%%%%%%%%%%%%%%%%%%%%%%%%%%%%%%%%

% Don't change these lines
\bsp	% typesetting comment
\label{lastpage}
\end{document}